\documentclass[prc,twocolumn,superscriptaddress,nofootinbib]{revtex4-1}

\usepackage{newtxtext}
\usepackage[varvw,bigdelims]{newtxmath}
\usepackage[colorlinks=true,allcolors=blue]{hyperref}
\usepackage{graphicx}
\usepackage{bm}
\usepackage{amsmath}

\begin{document}

\title{Collision integral with momentum-dependent potentials and its impact on pion production in heavy-ion collisions}

\author{Natsumi Ikeno}
\affiliation{Department of Agricultural, Life and Environmental Sciences, Tottori University, Tottori 680-8551, Japan}
\affiliation{RIKEN Nishina Center, 2-1 Hirosawa, Wako, Saitama 351-0198, Japan}
\affiliation{Cyclotron Institute, Texas A\&M University, College Station, Texas 77843, USA}
\author{Akira Ono}
\affiliation{Department of Physics, Tohoku University, Sendai 980-8578, Japan}
\affiliation{RIKEN Nishina Center, 2-1 Hirosawa, Wako, Saitama 351-0198, Japan}

\begin{abstract}
The momentum dependence of the nucleon mean-field potential in a wide momentum range can be an important factor to determine the $\Delta$ resonance and pion production in intermediate-energy heavy-ion collisions. In particular, in neutron-rich systems such as ${}^{132}\mathrm{Sn}+{}^{124}\mathrm{Sn}$ collisions, we need to carefully treat the momentum dependence because the neutron and proton potentials can have different momentum dependence, as characterized at low momenta by effective masses.
In the present work, we rigorously calculate the collision terms of $NN \leftrightarrow N \Delta$ and $\Delta \leftrightarrow N \pi$ processes with the precise conservation of energy and momentum under the presence of momentum-dependent potentials for the initial and final particles of the process.
The potentials affect not only the threshold condition for the process but also the cross section in general as a function of the momenta of the initial particles, which is treated in a natural way in the present work.
Calculations are performed by combining the nucleon dynamics obtained by the antisymmetrized molecular dynamics (AMD) model with a newly developed transport code which we call sJAM.
The calculated results for central ${}^{132}\mathrm{Sn}+{}^{124}\mathrm{Sn}$ collisions at 270 MeV/nucleon clearly show that the momentum dependence of the neutron and proton potentials has a significant impact on the $NN \to N \Delta$ process, and this information is strongly reflected in the charged pion ratio ($\pi^-/\pi^+$). 
We also investigate the effects of the high-density symmetry energy and the isovector part of the potential of $\Delta$ resonances on pion production, which we find are relatively small compared to the effect of the momentum dependence of the neutron and proton potentials.
\end{abstract}

\date{\today}

\maketitle
\section{Introduction}
Heavy-ion collisions provide useful systems for studying nuclear matter under various conditions of temperature and density. 
In particular, in collisions of neutron-rich nuclei, they allow us to access insights into the equation of state (EOS) of isospin-asymmetric nuclear matter at high density, where the numbers of protons and neutrons are unbalanced~\cite{horowitz2014,Russotto:2023ari}. The study of the EOS has recently attracted attention as a combined effort involving nuclear theory, experimental nuclear physics, and astrophysics. For example, by combining information from terrestrial and astrophysical observations, it has been reported that the properties of neutron-rich dense matter are constrained in the density range explored in neutron stars~\cite{Huth:2021bsp}. Constraints from heavy-ion collision experiments play an important role there.

In the incident energy range from several hundred MeV to several GeV per nucleon, experiments of the heavy-ion collisions have been carried out and information on EOS has been deduced to some extent from observables such as the collective flow and the kaon and pion production, e.g., by the analyses of the Au + Au collision data taken at GSI~\cite{Russotto:2023ari}.
Recently, at the RI Beam Factory (RIBF) in RIKEN, an experiment of collisions of Sn isotopes was performed at 270~MeV/nucleon by the S$\pi$RIT collaboration, which allows us to study the systems of various isospin asymmetries.
They reported 
 the pion observables~\cite{spirit2021PLB,spirit2021PRL} and the nucleon observables of the light fragments~\cite{SpRIT:2021dvt,Lee:2022,NishimuraSpRIT}.  
In particular, the charged pion ratio ($\pi^-/\pi^+$) is believed to be one of the good observables to probe the symmetry energy at high density~\cite{bali2002}.
The slope of the symmetry energy is determined to be $42 < L< 117$~MeV from the charged pion spectra at high transverse momenta~\cite{spirit2021PRL,Lynch:2021xkq} by an analysis with one of the transport models, dcQMD~\cite{cozma2021}.

Transport models are used as the main method to obtain physics information from heavy-ion collisions by solving the time evolution of the collision reactions~\cite{TMEP:2022}. However, there are some ambiguities in the model ingredients and numerical implementations.
The Transport Model Evaluation Project (TMEP) has been underway to resolve the uncertainties among the transport model predictions~\cite{spirit2021PLB,TMEP:2016tup,TMEP:2017mex,ono2019,TMEP:2021ljz,TMEP:2022,xu2023}. One of the projects was the pion production prediction of Ref.~\cite{spirit2021PLB}, where a significant discrepancy was found between the transport model predictions and the experimental data for the charged pion multiplicities and charged pion ratio in ${}^{132}\mathrm{Sn}+{}^{124}\mathrm{Sn}$, ${}^{112}\mathrm{Sn}+{}^{124}\mathrm{Sn}$, and ${}^{108}\mathrm{Sn}+{}^{112}\mathrm{Sn}$ collisions.
Most theoretical predictions including the AMD+JAM model~\cite{ikeno2016,ikeno2016erratum,Ikeno:2019mne} underestimated the $\pi^-/\pi^+$ ratio.
One of the reasons for this discrepancy is considered to be the lack of potentials for the nucleons and $\Delta$ resonances in the collision terms.

The momentum dependence of the neutron and proton mean-field potentials in isospin-asymmetric systems is one of the important aspect of the nuclear interaction that affects various phenomena in nuclear physics and astrophysics, see e.g. Refs.~\cite{bali2008,Li:2018lpy} for reviews. The nucleon collective flow in heavy-ion reactions is known to be strongly sensitive to the momentum dependence of the isoscalar part of the potential~\cite{danielewicz2000,Nara:2020ztb}, and also shown to be affected by the momentum dependence of the isovector part of the potential, i.e., the effective mass splitting at high isospin asymmetries~\cite{Rizzo:2003if,Li:2004zi}.
The $n/p$ spectral ratio is also predicted to be sensitive to the effective mass splitting~\cite{Zhang:2014sva} and some information has been obtained from experimental data~\cite{Coupland:2014gya,Morfouace:2019jky}.
Note that the above effects of the momentum-dependent potentials are essentially caused by the single-particle motion in the mean field.

The collision term in transport theory is as important as the mean-field propagation term. In fact, when the system reaches thermal equilibrium, the collision term determines the properties such as the EOS and the chemical composition. For example, careful treatments of the $NN\leftrightarrow N\Delta$ and $\Delta\leftrightarrow N\pi$ processes are necessary for a correct description of a mixture of nucleons, $\Delta$ resonances and pions in a box as shown in Ref.~\cite{ono2019}, where transport models were compared in the case without mean-field potentials. When potentials are present, the collision term needs to incorporate potentials at least to guarantee the correct description of equilibrium reflecting mean-field interaction. First of all, the presence of potentials affects the threshold condition for the process particularly when the potentials are momentum dependent and/or the particle species change from the initial to the final state, e.g., in the $NN\to N\Delta$ process. In an isospin-asymmetric environment where the neutron and proton potentials are different, the threshold condition depends on the isospin channel, e.g., whether $nn\to p\Delta^-$ or $pp\to n\Delta^{++}$, which requires an extended treatment in transport models. A few transport models consider the threshold effect in heavy-ion collision calculations~\cite{cozma2016,song2015} and its importance has  also been demonstrated in box calculations~\cite{ferini2005,zhenzhang2018,zhenzhang2018chi}. Furthermore, a related question is how the presence of potentials modifies the cross section above the threshold as a function of the momenta of the colliding particles, as investigated by Ref.~\cite{larionov2003} for isospin symmetric systems and by Ref.~\cite{cui2018} for asymmetric nuclear matter in the framework of  the one-boson exchange model. A fully consistent incorporation of the potential-dependent cross sections in transport calculations is still a challenging problem.

In the present work, we theoretically study the $\Delta$ resonance and pion productions in the ${}^{132}\mathrm{Sn}+{}^{124}\mathrm{Sn}$ collision at $E/A =270$~MeV by taking into account momentum-dependent mean-field potentials in the collision term. 
For this, we develop a transport model (AMD+sJAM) to properly treat the potentials, e.g., for the $NN \leftrightarrow N\Delta$ and $\Delta \leftrightarrow N \pi$ processes. This is an extension of the previous model AMD+JAM~\cite{ikeno2016,ikeno2016erratum,Ikeno:2019mne} by Ikeno, Ono, Nara, and Ohnishi in which the antisymmetrized molecular dynamics (AMD) \cite{ono1992} was combined with a hadronic cascade model (JAM) whose collision term was formulated for particles in vacuum \cite{nara1999}.
We will see that by the extension for potentials the results are improved drastically and the high $\pi^-/\pi^+$ ratio can be explained by the AMD+sJAM model, depending on the momentum dependence of the neutron and proton mean-field potentials.

This paper is organized as follows. 
In Sec.~\ref{sec:Pot}, we explain our choice of the nuclear interaction and the nucleon and $\Delta$ potentials.
In Sec.~\ref{sec:crosssection}, we formulate the collisions under the presence of potentials, especially for $NN \leftrightarrow N \Delta$ and $\Delta \leftrightarrow N \pi$ processes.
In Sec.~\ref{sec:nnnd}, as an example, we discuss how the $NN \rightarrow N \Delta$ cross sections in nuclear matter are affected by momentum-dependent potentials.
In Sec.~\ref{sec:AMDsJAM}, we introduce the AMD+sJAM transport model, in which the above formulation for the collision term is applied to a newly developed code sJAM.
In Sec.~\ref{sec:results}, we show the results of the pion observables in the ${}^{132}\mathrm{Sn}+{}^{124}\mathrm{Sn}$ collision at the incident energy of $E/A =270$~MeV within the AMD+sJAM model. 
We will see a strong impact of the momentum dependence of the neutron and proton potentials on the pion productions. 
We also investigate the effects of the high-density symmetry energy and the isovector part of the potential of the $\Delta$ resonances on the pion productions. 
A summary is given in Sec.~\ref{sec:summary}.

\section{Potentials}\label{sec:Pot}

\subsection{Energy density and potentials in system with nucleons only}\label{sec:Potamd}

When only nucleons are present in the system, our model is based on the interaction energy density expressed as
\begin{equation}
\begin{split}
\mathcal{E}_{\text{int}}(\bm{r}) = 
\sum_{\alpha \beta}  &\Bigl\{
U^{t_0}_{\alpha \beta} \rho_\alpha(\bm{r})  \rho_\beta (\bm{r}) 
 +
U^{t_3}_{\alpha \beta} 
 \rho_\alpha(\bm{r})  \rho_\beta (\bm{r}) [\rho(\mathbf{r})]^{\gamma}
\\
&+  U^\tau_{\alpha \beta} 
\tilde{\tau}_\alpha (\bm{r}) \rho_\beta (\bm{r})
+  U^{\nabla}_{\alpha \beta} \nabla \rho_\alpha(\bm{r})  \nabla \rho_\beta(\bm{r})
\Bigr\},
\end{split}
\label{eq:Edensity}
\end{equation}
which is similar to the Skyrme energy density functional but the spin--orbit term is not included in our calculations. Each single-particle state is assumed to be a product of the spatial part and the spin--isospin part $\chi_\alpha$, with the spin--isospin label $\alpha$ (or $\beta$) $=p\uparrow, p\downarrow, n\uparrow$ and $n\downarrow$. The densities $\rho_\alpha(\bm{r})$ and $\tilde{\tau}_\alpha(\bm{r})$ in Eq.~\eqref{eq:Edensity} are defined by using the one-body Wigner distribution function $f_\alpha(\bm{r},\bm{p})$ as
\begin{align}
\rho_\alpha(\bm{r})&=\int\frac{d\bm{p}}{(2\pi\hbar)^3}f_\alpha(\bm{r},\bm{p}),\\
\tilde{\tau}_\alpha(\bm{r})
&=\int\frac{d\bm{p}}{(2\pi\hbar)^3} \frac{[\bm{p}-\bar{\bm{p}}(\bm{r})]^2}
{1+[\bm{p}-\bar{\bm{p}}(\bm{r})]^2/\Lambda_{\text{md}}^2} f_\alpha(\bm{r},\bm{p}),
\end{align}
with
\begin{equation}
\bar{\bm{p}}(\bm{r})=\frac{1}{\sum_\alpha\rho_\alpha(\bm{r})}
\sum_\alpha\int\frac{d\bm{p}}{(2\pi\hbar)^3}
\bm{p}f_\alpha(\bm{r},\bm{p}).
\end{equation}
Here $\tilde{\tau}_\alpha(\bm{r})$ is a kind of the kinetic energy density but a cut-off parameter $\Lambda_{\text{md}}$ has been introduced in Ref.~\cite{ikeno2016} following Ref.~\cite{Gale:1987zz}, which will be important for the high-momentum behavior of the mean field.

The coefficients $U^{t_0}_{\alpha \beta} $, $U^{t_3}_{\alpha \beta}$, $U^\tau_{\alpha \beta}$, and $U^{\nabla}_{\alpha \beta}$ in Eq.~\eqref{eq:Edensity} are related to the Skyrme parameters by
\begin{align}
U^{t_0}_{\alpha\beta}&= \frac{1}{2}t_0 \langle \alpha \beta | (1+x_0P_\sigma) | \alpha \beta - \beta \alpha \rangle,\\
U^{t_3}_{\alpha\beta}&= \frac{1}{12} t_3 \langle \alpha \beta |  (1+x_3P_\sigma) | \alpha \beta - \beta \alpha \rangle,\\
U^{\tau}_{\alpha\beta}&= \frac{1}{4} t_1 \langle \alpha \beta | (1+x_1P_\sigma) | \alpha \beta - \beta \alpha \rangle  \nonumber \\
&\qquad + \frac{1}{4}t_2 \langle \alpha \beta |(1+x_2P_\sigma)~ | \alpha \beta  + \beta \alpha \rangle,  \\
U^{\nabla}_{\alpha\beta}&= \frac{3}{16}t_1 \langle \alpha \beta | (1+x_1P_\sigma) | \alpha \beta - \beta \alpha \rangle \nonumber \\
&\qquad  - \frac{1}{16}t_2 \langle \alpha \beta |(1+x_2P_\sigma)~ | \alpha \beta  + \beta \alpha \rangle ,
\end{align}
where $P_\sigma$ is the spin exchange operator.
In the case of $\Lambda_{\text{md}}=\infty$, our interaction is equivalent to the Skyrme-type interaction
\begin{equation}
\begin{split}
v_{ij}&=t_0(1+x_0P_\sigma)\delta(\mathbf{r})\\
 &\quad+\tfrac{1}{2}t_1(1+x_1P_\sigma)
[\delta(\mathbf{r})\mathbf{k}^2 +\mathbf{k}^2\delta(\mathbf{r})]\\
&\quad+t_2(1+x_2P_\sigma)
\mathbf{k}\cdot\delta(\mathbf{r})\mathbf{k}\\
&\quad+\tfrac{1}{6}{t_3}(1+x_3P_\sigma)[\rho(\mathbf{r}_i)]^{\gamma}\delta(\mathbf{r}), \end{split}
\label{eq:eff_int}
\end{equation}
where $\mathbf{r}=\mathbf{r}_i-\mathbf{r}_j$ and
$\mathbf{k}=\frac{1}{2\hbar}(\mathbf{p}_i-\mathbf{p}_j)$.

Employing the AMD model \cite{ono1992}, we solve the time evolution of a many-nucleon system by directly using the energy density functional [Eq.~\eqref{eq:Edensity}] together with the other parameters, e.g., for the two-nucleon collision term. However, for some purposes, it is useful to consider the corresponding momentum-dependent mean-field potential which is obtained by
\begin{equation}
U_\alpha(\bm{r},\bm{p})=(2\pi\hbar)^3\frac{\delta}{\delta f_\alpha(\bm{r},\bm{p})}
\int\mathcal{E}_{\text{int}}(\bm{r})d\bm{r}.
\label{eq:upot_def}
\end{equation}
In the case of Eq.~\eqref{eq:Edensity}, we have
\begin{equation}
U_\alpha(\bm{r},\bm{p})
=A_\alpha(\bm{r})\frac{[\bm{p}-\bar{\bm{p}}(\bm{r})]^2}
{1+[\bm{p}-\bar{\bm{p}}(\bm{r})]^2/\Lambda_{\text{md}}^2}
+\tilde{C}_\alpha(\bm{r}),
\label{eq:upot}
\end{equation}
with
\begin{equation}
A_\alpha(\bm{r})=\sum_\beta U^\tau_{\alpha\beta}\rho_\beta(\bm{r})
\end{equation}
and
\begin{equation}
\begin{split}
\tilde{C}_\alpha(\bm{r})&=
\sum_{\beta} \Bigl\{
2U^{t_0}_{\alpha \beta}\rho_\beta(\bm{r})
 + 2 U^{t_3}_{\alpha \beta} \rho_\beta(\bm{r}) [\rho(\mathbf{r})]^\gamma\\
&\qquad\qquad + U^\tau_{\alpha \beta}\tilde{\tau}_\beta(\bm{r})
- 2 U^{\nabla}_{\alpha \beta}   \nabla^2 \rho_\beta(\bm{r})
  \Bigr\}
\\ 
&\quad +
 \biggl( \sum_{\alpha' \beta'} U^{t_3}_{\alpha' \beta'} 
 \rho_{\alpha'}(\bm{r}) \rho_{\beta'} (\bm{r})  \biggr)
 \gamma [\rho(\mathbf{r})]^{\gamma-1}.
\end{split}
\label{eq:Ctilde}
\end{equation}
In the above, the term originating from $\partial\tilde{\tau}_\alpha(\bm{r})/\partial\bar{\bm{p}}(\bm{r})$ has been ignored, which is justified when
\begin{equation}
\int\frac{d\bm{p}'}{(2\pi\hbar)^3}
\frac{\bm{p}'-\bar{\bm{p}}(\bm{r})}
{[1+(\bm{p}'-\bar{\bm{p}}(\bm{r}))^2/\Lambda_{\text{md}}^2]^2}
f_\alpha(\bm{r},\bm{p}')\approx 0.
\end{equation}

\begin{figure}
\centering
\includegraphics[width=\columnwidth]{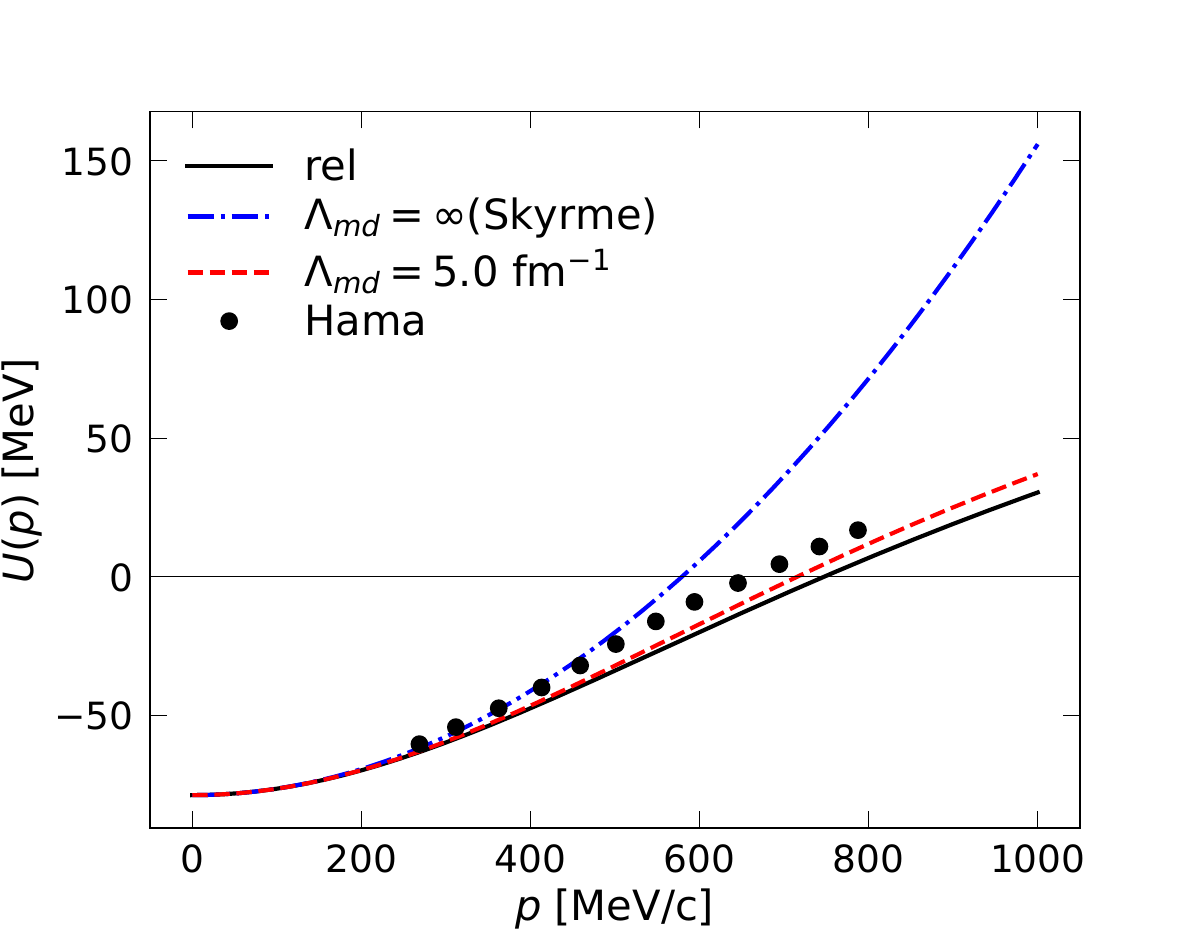}
\caption{The momentum dependence of the nucleon potentials in the symmetric nuclear matter ($\rho_n=\rho_p$) at the normal density $\rho=0.16$~fm$^{-3}$ and at zero temperature. 
The blue dot--dashed line shows the nucleon potential with $\Lambda_{\text{md}}=\infty$ (Skyrme interaction of Eq.~\eqref{eq:eff_int}), the red dashed line the one with $\Lambda_{\text{md}}/\hbar= 5$~fm$^{-1}$ in Eq.~\eqref{eq:upot},  
and the black solid line the one of the relativistic form in Eq.~\eqref{eq:Urel}.
The solid points indicate the empirical optical potential from Hama et al.~\cite{Hama:1990vr}. }
\label{fig:pot_mdcorr}
\end{figure}

An example of the momentum dependent potential $U(p)=U_\alpha(\bm{r},\bm{p})$ is shown in Fig.~\ref{fig:pot_mdcorr} for the zero-temperature symmetric nuclear matter at the saturation density $\rho=\rho_0=0.16\ \text{fm}^{-3}$. Here, we took the SLy4 parameter set \cite{Chabanat:1997un} to determine the coefficients in Eq.~\eqref{eq:Edensity}. The blue dot--dashed curve shows the case of $\Lambda_{\text{md}}=\infty$, in which the momentum dependence is simply quadratic in $p$ as a direct consequence of the Skyrme interaction of Eq.~\eqref{eq:eff_int}, and the curvature is related to the effective nucleon mass $m^*\approx 0.70\,m_N$ in the SLy4 parametrization. This quadratic momentum dependence is too strong compared to the solid points which show the energy or momentum dependence of the optical potential derived from the global fit of the proton--nucleus elastic scattering data by Hama et al.~\cite{Hama:1990vr}. On the other hand, when we choose the parameter $\Lambda_{\text{md}}/\hbar=5.0\ \text{fm}^{-1}$, the momentum dependence of $U(p)$ is weakened as shown by the red dashed line in Fig.~\ref{fig:pot_mdcorr}, and it is now similar to the potential by Hama. For the present study of heavy-ion collisions at 270 MeV/nucleon, in particular for the production of $\Delta$ resonances and pions, we expect that a suitable description of $U(p)$ at $p\gtrsim 500$ MeV/$c$ is important. Therefore, in all calculations in this paper, we take $\Lambda_{\text{md}}/\hbar=5.0\ \text{fm}^{-1}$.

To formulate the collision term in Sec.~\ref{sec:crosssection}, we use a relativistic framework in which the nucleon single-particle energy is written with the scalar and vector potentials (self-energies) $\Sigma_a(\bm{r}) \equiv (\Sigma_a^{s}(\bm{r}), \Sigma_a^{0}(\bm{r}), \bm{\Sigma}_a(\bm{r}))$ as
\begin{equation}
 E_a(\bm{r, p}) = \sqrt{(m_N + \Sigma_a^s(\bm{r}))^2 + (\bm{p}- \bm{\Sigma}_a(\bm{r}))^2 } + \Sigma_a^0 (\bm{r}).
\label{Esjam} 
\end{equation}
We assume here that the distribution does not depend on the direction of the spin, and thus the index $a$ takes $p$ (proton) and $n$ (neutron). The relations such as $f_p=f_{p\uparrow}+f_{p\downarrow}$ and $U_p=U_{p\uparrow}=U_{p\downarrow}$ should be understood implicitly. The scalar and vector potentials $\Sigma_a(\bm{r})$ can be determined from the potential $U_a(\bm{r},\bm{p})$ of Eq.~\eqref{eq:upot} by following the same prescription used in Ref.~\cite{zhenzhang2018chi}. We require the equivalence
\begin{multline}
\frac{\bm{p}^2}{2m_N}+A_a(\bm{p}-\bar{\bm{p}})^2+\tilde{C}_a +m_N \\
\approx
\sqrt{(m_N+\Sigma_a^s)^2+(\bm{p}-\bm{\Sigma}_a)^2}+\Sigma_a^0
\end{multline}
to hold at low momenta up to the order of $(\bm{p}-\bm{\Sigma}_a)^2$. From this condition, the scalar potential is determined by
\begin{equation}
\Sigma^s_a=m^*_a - m_N
\end{equation}
with the nucleon effective mass
\begin{equation}
m^*_a= (m_N^{-1} + 2A_a)^{-1},
\end{equation}
and the vector potential is derived as
\begin{align}
\bm{\Sigma}_a&=4A_a m^*_a\bar{\bm{p}},\\
\Sigma^0_a&=\tilde{C}_a-\Sigma^s_a +A_a\bar{\bm{p}}^2
-8m^*_a A_a^2\bar{\bm{p}}^2.
\end{align}
However, when $\bar{\bm{p}}\rho_b$ can be identified with the current $\bm{J}_b$ defined by
\begin{equation}
\bm{J}_b(\bm{r})=\int\frac{d\bm{p}}{(2\pi\hbar)^3}\bm{p}f_b(\bm{r},\bm{p}),
\end{equation}
we may use an alternative formula for the vector potential
\begin{align}
\bm{\Sigma}_a&= 2m^*_a\sum_b U^\tau_{ab}\bm{J}_b,\\
\Sigma^0_a&= C_a-\Sigma^s_a-\frac{\bm{\Sigma}_a^2}{2m^*_a},
\end{align}
where $C_a$ is the same as $\tilde{C}_\alpha$ but $\tilde{\tau}_\beta(\bm{r})$ in Eq.~\eqref{eq:Ctilde} is replaced by the so-called kinetic energy density
\begin{equation}
\tau_b(\bm{r})=\int\frac{d\bm{p}}{(2\pi\hbar)^3}\bm{p}^2f_b(\bm{r},\bm{p}).
\end{equation}

In Fig.~\ref{fig:pot_mdcorr}, the red dashed curve shows a relativistic version of the potential
\begin{equation}
U_{\text{rel}}(p)=\sqrt{(m_N+\Sigma^s)^2+p^2}+\Sigma^0-\sqrt{m_N^2+p^2}
\label{eq:Urel}
\end{equation}
for the zero-temperature symmetric nuclear matter at $\rho=\rho_0$. The scalar and vector potentials are determined from the SLy4 parameters as described above. We find that the momentum dependence of $U_{\text{rel}}(p)$ is similar to the empirical optical potential by Hama et al.~\cite{Hama:1990vr} (solid points) and also consistent with the nonrelativistic one with $\Lambda_{\text{md}}/\hbar=5.0~\text{fm}^{-1}$ (red dashed curve).

\begin{figure*}
\centering
\includegraphics[scale=0.35]{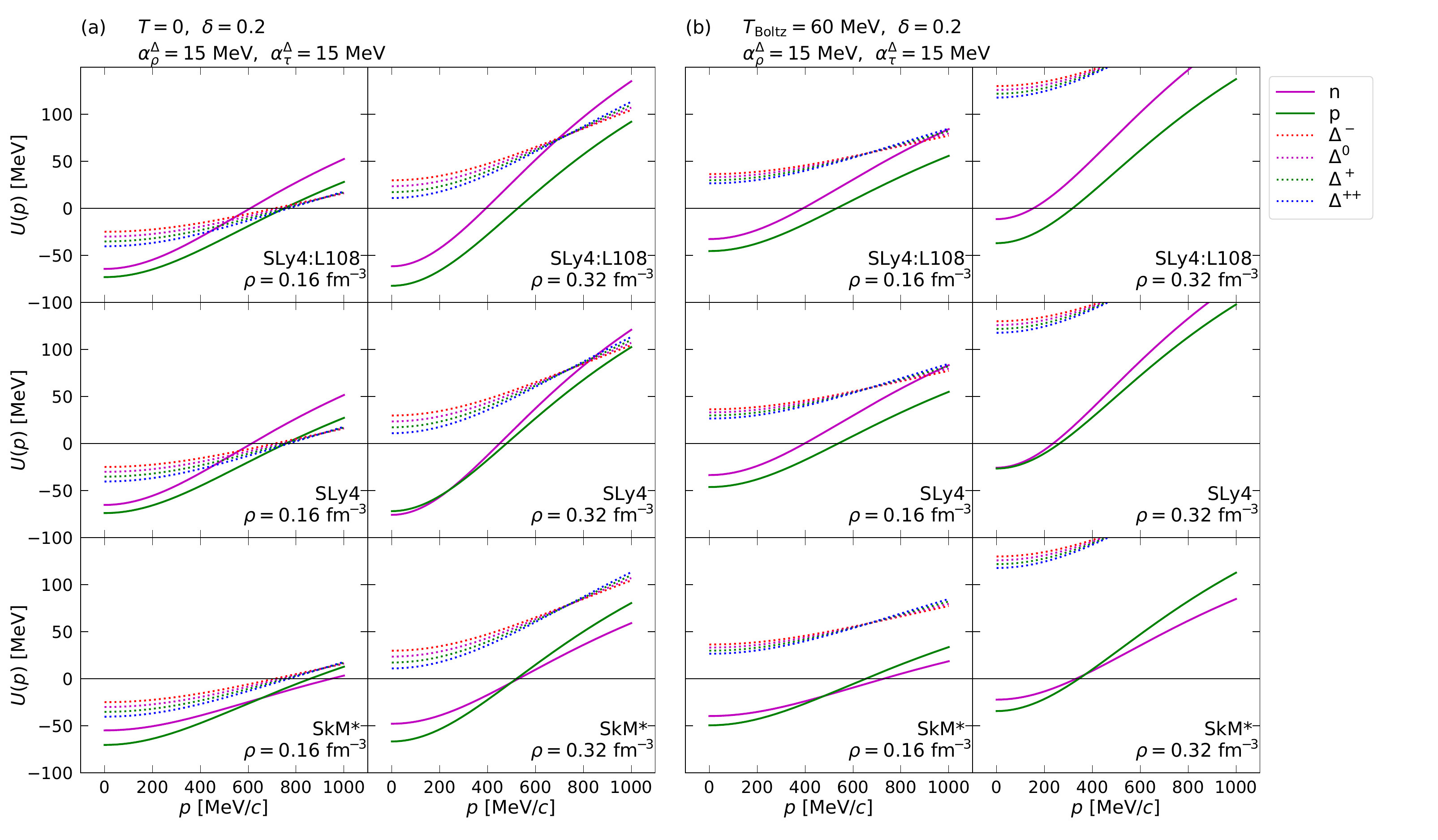}
\caption{ Left part (a): The momentum dependence of the potentials of the nucleons and $\Delta$ resonances in the asymmetric nuclear matter $\delta = (\rho_n - \rho_p)/(\rho_n + \rho_p) = 0.20$ at zero temperature $T = 0$. The left column is for the density $\rho=0.16\ \text{fm}^{-3}$ and the right column is for $\rho=0.32\ \text{fm}^{-3}$. The solid lines in each panel indicate the neutron and proton potentials in the relativistic form, for the three cases of parametrizations based on SLy4:L108 (top), SLy4 (middle), and SkM* (bottom).
The dotted lines show the potentials of $\Delta$ resonances ($\Delta^-, \Delta^0, \Delta^+, \Delta^{++}$) adopted in the present work. As explained in Sec.~\ref{sec:Dpot}, the $\Delta$ potentials are based on the SkM* parametrization with additional repulsive terms with parameters $\alpha_\rho^\Delta = 15$~MeV and $\alpha_\tau^\Delta = 15$~MeV. The isospin splitting is controlled by a parameter $\gamma^\Delta = 1$.
Right part (b): The same as (a) except at a finite temperature corresponding to the kinetic energy density $\tau_b=3m_NT_{\text{Boltz}}\rho_b$ with $T_{\rm Boltz} = 60$ MeV.}
 \label{fig:pot}
\end{figure*}

In the present study, we investigate how the results depend on the effective interaction by comparing three cases of the energy density functional which are labeled as `SLy4', `SLy4:L108', and `SkM*'. In all cases, we set $\Lambda_{\text{md}}/\hbar=5.0\ \text{fm}^{-1}$ to modify the momentum dependence. 
The `SLy4' functional is based on the Skyrme SLy4 force of Ref.~\cite{Chabanat:1997un}, for which the corresponding nuclear-matter incompressibility is $K=230$~MeV at the saturation density $\rho_0=0.160\ \text{fm}^{-3}$, as summarized in Table~\ref{tab:1} together with other nuclear matter properties.
The nuclear symmetry energy at the saturation density $\rho_0$ is $S_0=32.0$~MeV with the slope parameter $L=46$~MeV (called `asy-soft' or soft symmetry energy). 
The `SLy4:L108' functional is based on a Skyrme parameter set which is obtained by modifying the $x_3$ and $x_0$ parameters of the SLy4 interaction to have a stiff symmetry energy with $L=108$~MeV \cite{ikeno2016} (called `asy-stiff' or stiff symmetry energy) without changing $S_0$ and the properties of the symmetric nuclear matter. The `SkM*' functional is based on the SkM* parameter set of Ref.~\cite{Bartel:1982ed}, which corresponds to $K=217$~MeV, $S_0=30.0$~MeV and $L=46$ (`asy-soft') of the nuclear matter at the saturation density $\rho_0=0.160\ \text{fm}^{-3}$. In the symmetric nuclear matter at $\rho_0$, nucleons have an effective mass $m^*=0.70 \, m_N$ for SLy4 and SLy4:L108, and $m^*=0.79\, m_N$ for SkM*.

\begin{table}[tb]
 \centering
\caption{ Nuclear matter properties for the effective interactions of Skyrme SLy4~\cite{Chabanat:1997un}, SLy4:L108~\cite{ikeno2016}, and SkM*~\cite{Bartel:1982ed} (see text). 
\label{tab:1}
} 
\begin{tabular}{cccc}
\hline  &~~SLy4~~ &~~SLy4:L108~~&~~SkM*~~ \\
\hline
$\rho_0$ [fm$^{-3}$] & 0.160  &  0.160 & 0.160   \\
$E/A$ [MeV] & $-15.97$ & $-15.97$ & $-15.77$    \\
$K$ [MeV]& 230 & 230 & 217 \\
$m^*/m_N$ & 0.70 & 0.70  & 0.79 \\
$S_0$ [MeV] & 32.0 & 32.0 & 30.0   \\
$L$ [MeV] & 46 & 108 & 46  \\
$ \Delta m^*_{np}/(m_N \delta)$~~~& $-0.18$ & $-0.18$ &  $+0.33$   \\
in n-rich & $m^*_n < m^*_p$ & $m_n^*<m_p^*$ & $m_n^*>m_p^*$   \\
\hline
\end{tabular}
\end{table}

By the solid lines in Fig.~\ref{fig:pot}, we show the neutron and proton potentials $U_n(p)$ and $U_p(p)$ as a function of the momentum, after converting them to the relativistic form of Eq.~\eqref{eq:Urel}. (The $\Delta$ potentials shown by the dashed lines are to be discussed later in Sec.~\ref{sec:Dpot}.) For the zero-temperature asymmetric nuclear matter with $\delta=(\rho_n-\rho_p)/(\rho_n+\rho_p)=0.2$, the potentials based on SLy4:L108 (top), SLy4 (middle) and SkM* (bottom) are shown at the density $\rho=\rho_0$ (left) and $\rho=2\rho_0$ (right) in the left part (a) of Fig.~\ref{fig:pot}. Evidently, in these asymmetric cases, the momentum dependence of $U_n$ is different from that of $U_p$. At $\rho_0$, the neutron--proton effective mass difference is $\Delta m^*_{np}=m^*_n-m^*_p= -34.4~\text{MeV}= -0.18 \, m_N \delta $ for both SLy4 and SLy4:L108, while it is $\Delta m^*_{np}= 61.5~\text{MeV} = 0.33 \,m_N \delta$ for SkM*. It should be noted that $U_n$ and $U_p$ at $\rho=\rho_0$ for SLy4 are identical to $U_n$ and $U_p$ for SLy4:L108, respectively. When the density is raised to $2\rho_0$ (right panels) where the symmetry energy is larger in SLy4:L108 than in SLy4, the neutron potential $U_n$ (or the proton potential $U_p$) in SLy4:L108 is shifted upwards (or downwards) compared to that in SLy4. Thus, the gap between $U_n$ and $U_p$ is related to the symmetry energy. In the SkM* case, the momentum dependence of $U_n$ is weak compared to that of $U_p$, i.e.\ $m^*_n>m^*_p$, and consequently $U_n(p)$ becomes even lower than $U_p(p)$ at high momenta, $p>650$ MeV/$c$ for $\rho_0$ and $p >500$ MeV/$c$ for $2\rho_0$. We will see later in Sec.~\ref{sec:results} how these behaviors of $U_n$ and $U_p$ affect the nucleon dynamics in heavy-ion collisions and the production of $\Delta$ resonances.

In the right part (b) of Fig.~\ref{fig:pot}, the neutron and proton potentials are shown for a high-temperature nuclear matter, in the same way as in Fig.~\ref{fig:pot} (a) for the $T=0$ case. Here the potentials are shown for the kinetic energy density $\tau_b=3m_NT_{\text{Boltz}}\rho_b$ with $T_{\text{Boltz}}=60$ MeV, which may be close to the situation in heavy-ion collisions studied in this paper. The potentials are generally higher than in the $T=0$ case because of their $\tau_b$ dependence. However, the qualitative behaviors observed at $T=0$ are preserved even at this high temperature.

\subsection{$\Delta$ potentials\label{sec:Dpot}}
The potential for the $\Delta$ resonance in nuclei has been studied by the theoretical analyses of the pion-nucleus, photon-nucleus, and electron-nucleus scatterings for decades (see, e.g., Ref.~\cite{Mosel:2020zdw}).
For example, the potential depth of the $\Delta$ resonance was reported to be about $-30$~MeV in a nucleus~\cite{Hirata:1978wp,Freedman:1981dz}, and later,
 to be about $-23 \, \rho/\rho_0$ MeV~\cite{Oset:1987re} and $-33 \,\rho/\rho_0$ MeV \cite{Garcia-Recio:1989hyf} where $\rho_0$ is the central density. 
All these potentials for the $\Delta$ resonance in nuclei show less binding compared to nucleon potentials.
On the other hand, there is still some room for debate on that topic~\cite{Bodek:2020wbk}. 
Also, the $\Delta$ potential seems to play an important role in the neutron star studies \cite{Drago:2014oja,Kolomeitsev:2016ptu,Raduta:2021xiz,Marquez:2022fzh}.

In transport model calculations for heavy-ion collisions in the literature, the $\Delta$ potentials $U_\Delta$ (or $\Sigma_\Delta$) are often linked with the nucleon potentials $U_N$ (or $\Sigma_N$) by linear combinations such as $U_{\Delta^-}=U_n$, $U_{\Delta^0}=\frac23U_n+\frac13U_p$, $U_{\Delta^+}=\frac13U_n+\frac23U_p$, and $U_{\Delta^{++}}=U_p$ \cite{bali2008,zhenzhang2018chi,ferini2005,cozma2016}. In this case, when one varies $U_N$ to study the sensitivity to the nuclear interaction such as the symmetry energy, observables will change not only through the change of $U_N$ but also through that of $U_\Delta$. Then the results have to be interpreted carefully, considering that the link of $U_\Delta$ to e.g.\ the nuclear symmetry energy has not been established theoretically. In contrast, in the present work, we treat $U_N$ and $U_\Delta$ as independent variables, i.e., we do not change the parameters for $U_\Delta$ when $U_N$ is varied, similarly to the work by Cozma el al.~\cite{cozma2021,spirit2021PRL}.

We write the single-particle energy of a $\Delta$ resonance in a relativistic form
\begin{equation}
E(\bm{r},\bm{p})=\sqrt{(m_\Delta+\Sigma^s_\Delta(\bm{r}))^2+(\bm{p}-\bm{\Sigma}_\Delta(\bm{r}))^2}+\Sigma^0_\Delta(\bm{r})
\end{equation}
by using the scalar and vector potentials $(\Sigma_\Delta^{s}, \Sigma_\Delta^{0}, \bm{\Sigma}_\Delta)$. The vacuum mass $m_\Delta$ is distributed according to the spectral function of the resonance such as of a Breit--Wigner form.
In this article, we treat each component of the potential $\Sigma_\Delta = (\Sigma_\Delta^{s}, \Sigma_\Delta^{0}, \bm{\Sigma}_\Delta)$ as consisting of the isoscalar part $ \Sigma_{\rm is}$ and the isovector part $\Sigma_{\rm iv}$ as
\begin{equation}
\label{eq:Sigma_Delta}
\begin{split}
\Sigma_{\Delta^{-}} &=  \Sigma_{\rm is} +  \tfrac{3}{2} \Sigma_{\rm iv},\\
\Sigma_{\Delta^{0}} &= \Sigma_{\rm is} +  \tfrac{1}{2} \Sigma_{\rm iv},\\
\Sigma_{\Delta^{+}} &= \Sigma_{\rm is} -  \tfrac{1}{2} \Sigma_{\rm iv},\\
\Sigma_{\Delta^{++}} &= \Sigma_{\rm is} - \tfrac{3}{2} \Sigma_{\rm iv}.
\end{split}
\end{equation}
The isoscalar part $\Sigma_{\rm is} = (\Sigma_{\rm is}^{s}, \Sigma_{\rm is}^{0}, \bm{\Sigma}_{\rm is})$ is chosen as
\begin{equation}
\label{eq:Uis_Delta}
\begin{split}
\Sigma^{s}_{\rm is} &= \frac{1}{2}(\Sigma_n^s + \Sigma_p^s)_{\text{SkM*}},\\
\Sigma^{0}_{\rm is} &= \frac{1}{2} (\Sigma_n^0 + \Sigma_p^0)_{\text{SkM*}} + \alpha_\rho^\Delta  \frac{\rho}{\rho_0}  + \alpha_\tau^\Delta  \frac{\tau}{\tau_0} , \\
\bm{\Sigma}_{\rm is} &=  \alpha_\rho^\Delta  \frac{\bm{J}}{\rho_0},
\end{split}
\end{equation}
which is based on the nucleon potential in the SkM* parametrization, regardless of the actual nucleon potential that we choose from SLy4, SLy4:L108 and SkM* as explained in Sec.~\ref{sec:Potamd}. Assuming that the $\Delta$ potential is less attractive than the nucleon potential, we add repulsive terms in $\Sigma^0_{\text{is}}$ that linearly depends on the density $\rho=\rho_{p\uparrow}+\rho_{p\downarrow}+\rho_{n\uparrow}+\rho_{n\downarrow}$ and the kinetic energy density $\tau=\tau_{p\uparrow}+\tau_{p\downarrow}+\tau_{n\uparrow}+\tau_{n\downarrow}$. A corresponding term is included in $\bm{\Sigma}_{\text{is}}$ with
$\bm{J} = \bm{J}_{p\uparrow} + \bm{J}_{p\downarrow} + \bm{J}_{n\uparrow} + \bm{J}_{n\downarrow}$.
We will adjust the parameters $\alpha_\rho^\Delta$ and $\alpha_\tau^\Delta$ to reproduce the experimental data for the overall pion multiplicity in Sec.~\ref{sec:results}. The kinetic energy density is normalized by $\tau_0=\frac35p_{\text{F}}^2\rho_0$ with $p_{\text{F}}= (\frac{3}{2} \pi^2 \rho_0)^{1/3}$.

As for the isovector part of the $\Delta$ potential $\Sigma_{\rm iv} = (\Sigma_{\rm iv}^{s}, \Sigma_{\rm iv}^{0}, \bm{\Sigma}_{\rm iv})$, we use the neutron--proton potential difference in the SkM* parametrization as
\begin{align}
\label{eq:Uiv_Delta}
\Sigma^s_{\rm iv} &= \tfrac13\gamma^\Delta (\Sigma_n^s - \Sigma_p^s)_{\text{SkM*}} ,\\
\Sigma^0_{\rm iv} &= \tfrac13\gamma^\Delta (\Sigma_n^0 - \Sigma_p^0)_{\text{SkM*}} , \\
\bm{\Sigma}_{\rm iv} &= \bm{0},
\end{align}
where we have introduced a parameter $\gamma^\Delta$ to vary the isospin splitting of the $\Delta$ potentials. The case of $\gamma^\Delta =1$ corresponds to the relation chosen by Refs.~\cite{bali2008,zhenzhang2018chi,ferini2005,cozma2016} in which $\Sigma_{\Delta^{-}} - \Sigma_{\Delta^{++}} = \Sigma_n -  \Sigma_p$. On the other hand, the case of $\gamma^\Delta =3$ is another option with a large isospin splitting taken by Refs.~\cite{UmaMaheswari:1997mc,hong2014} in which $\Sigma_{\Delta^{-}} - \Sigma_{\Delta^{++}} = 3(\Sigma_n -  \Sigma_p)$.

The $\Delta$ potentials in asymmetric nuclear matter are indicated in Fig.~\ref{fig:pot} with the dotted lines, for the choice of the parameters $\alpha^\Delta_\rho=15$ MeV, $\alpha^\Delta_\tau=15$ MeV and $\gamma^\Delta=1$. 
In our study, we use this parameter choice as the default setting when we investigate how the momentum dependence of the nucleon potential affects the pion observables.

\section{Cross section under potentials\label{sec:crosssection}}

In this section, we formulate the cross sections and the resonance decay rates under the presence of potentials. In particular, the inelastic processes $NN\leftrightarrow N\Delta$ and $\Delta\leftrightarrow N\pi$ are important in the present study. In general, let us consider the cross section $\sigma(\bm{p}_1,\bm{p}_2)$ of a reaction channel which is a function of the canonical momenta $\bm{p}_1$ and $\bm{p}_2$ of the particles in the initial state. This function may be modified in a medium for three reasons. First, the matrix element of the process may change in the medium through the modification of intermediate states. Second, the phase space factor for the final state will change mainly because the momenta of the final particles depend on the potentials through energy conservation. In particular, the potentials affect the condition on $\bm{p}_1$ and $\bm{p}_2$ for the process to be energetically possible, which is often called the threshold effect in the literature \cite{ferini2005,song2015,cozma2016,zhenzhang2018,zhenzhang2018chi}. Third, the cross section includes the inverse of the flux of the initial particles, which is also affected by the momentum dependence of the potentials through the dispersion relation. In the present work, we carefully treat the last two sources of the potential effect in $\sigma(\bm{p}_1,\bm{p}_2)$. In heavy-ion collision calculations, we will calculate the cross section at every chance of collisions using the current values of potentials of the initial and final particles.
In this section, we use the natural units, $\hbar=c=1$.

\subsection{General binary process}

First, we consider a scattering process $1 + 2 \to 3 + 4$ occurring around a point $\bm{r}$ at a time $t$ in a heavy-ion collision. This process is assumed to change the momenta of the participating two particles under the constraint of the energy and momentum conservation. 
We generally consider an inelastic process in which the particle species (3 and 4) in the final state may be different from those (1 and 2) in the initial state.
A resonance is treated by randomly assigning the mass $m$ according to the spectral function $A(m)$. We here consider the case in which the particle 4 is a resonance. The case of a stable particle with a mass $M$ corresponds to a $\delta$ function as the spectral function, i.e., $A(m) = (2\pi) \delta(m-M)$.

The transition probability per unit time and unit volume, to produce a resonance particle with a mass between $m_4$ and $m_4 + dm_4$, can be expressed by using a Lorentz invariant matrix element as
\begin{equation}
\begin{split}
\frac{dW}{dVdt}&=\frac{A(m_4)dm_4}{2\pi}
\frac{d^3\bm{p}_3}{(2\pi)^32E_3^*} \frac{d^3\bm{p}_4}{(2\pi)^32E_4^*}
 |\langle\bm{p}_3\bm{p}_4|\mathcal{M}|\bm{p}_1 \bm{p}_2 \rangle_\Sigma|^2 \\
&\quad\times 2\pi\delta(E_3 + E_4- E_1- E_2)\\
&\quad\times (2\pi)^3\delta^3(\bm{p}_3 + \bm{p}_4 - \bm{p}_1 - \bm{p}_2),
\end{split}
\end{equation}
where the energies of the particles $i = 1,2,3$ and 4 depend on the scalar and vector potentials $\Sigma _i= (\Sigma_{i}^s, \Sigma_{i}^0, \bm{\Sigma}_i)$ as 
\begin{equation} 
E_i = E_i^* +\Sigma_{i}^0,
\end{equation}
with the effective mass and the kinetic energy and momentum
\begin{align}
m^*_i &= m_i +\Sigma_{i}^s,\\
E_i^*&= \sqrt{m_i^{*2}+\bm{p}^{*2}_i},\\
\bm{p}_i^*&=\bm{p}_i -\bm{\Sigma}_i .
\end{align}
For this invariant transition rate, it is convenient to perform the integration over the final momenta in the `out' frame which is defined by the Lorentz transformation with the velocity~\cite{song2015,zhenzhang2018}
\begin{equation}
\bm{\beta}_{\text{out}}=
\frac{\bm{p}^*_1+\bm{p}^*_2 + \bm{\Sigma}_1 + \bm{\Sigma}_2 -\bm{\Sigma}_3-\bm{\Sigma}_4 }{E^*_1+E^*_2 + \Sigma_{1}^0 + \Sigma_{2}^0 -\Sigma_{3}^0 -\Sigma_{4}^0 }.
\end{equation}
After integrating the transition probability over $\bm{p}_4$ and changing the integration variable from $\bm{p}_3$ to $\bm{p}^*_3 = (p^*_{\text{f}}, \Omega^*_{\text{f}})$, we obtain
\begin{equation}
\begin{split}
\frac{dW}{dVdt} &= \int 
\frac{p^{*2}_{\text{f}} dp^{*}_{\text{f}} d\Omega_f^*}{(2\pi)^3 4 E_3^* E_4^*}
|\mathcal{M}|_\Sigma^2 
\frac{A(m_4)dm_4}{2\pi} \\
&\quad\times 2\pi \delta\bigl(E_3(p^*_{\text{f}}) + E_4(p^*_{\text{f}}) - E_1- E_2 \bigr),
\end{split}
\end{equation}
and thus
\begin{equation}
\frac{dW}{dVdt}=
\frac{|\mathcal{M}|_\Sigma^2}{\pi}
\biggl[
\frac{p_{\text{f}}^{*2}}{4 v_{\text{f}} E_3^* E_4^*}
\biggr]_{\text{out}}
\frac{A(m_4)dm_4}{2\pi} \frac{d\Omega_{\text{f}}^*}{4\pi}
\end{equation}
where the quantity in $[...]_{\rm out}$ needs to be evaluated in the `out' frame, and $v_{\rm f}$ is the relative velocity of the final particles in that frame,
\begin{equation}
  v_{\text{f}} =\frac{p_{\text{f}}^*}{E_3^*}+\frac{p_{\text{f}}^*}{E_4^*}.
\end{equation}
Dividing the transition rate $dW/dVdt$ by the flux times the density $  v_{\text{i}} (2E^*_1) (2E^*_2)$
 of the initial state in the `in' frame defined by the velocity
\begin{equation}
\bm{\beta}_{\text{in}}=
\frac{\bm{p}^*_1+\bm{p}^*_2 }{E^*_1+E^*_2},
\end{equation}
the cross section is written as
\begin{equation}
d\sigma = \frac{|\mathcal{M}|_\Sigma^2}{\pi}
\biggl[
\frac{1}{4 v_{\text{i}} E_1^* E_2^*}
\biggr]_{\text{in}}
\biggl[
\frac{p_{\text{f}}^{*2}}{4 v_{\text{f}} E_3^* E_4^*}
\biggr]_{\text{out}}
\frac{A(m_4)dm_4}{2\pi} \frac{d\Omega_{\text{f}}^*}{4\pi}
\end{equation}
with the initial relative velocity
\begin{equation}
 v_{\text{i}} =\frac{p_{\text{i}}^*}{E_1^*}+\frac{p_{\text{i}}^*}{E_2^*}
\end{equation}
in the `in' frame where $\bm{p}^*_1 = -\bm{p}^*_2$ and $|\bm{p}^*_1| = |\bm{p}^*_2| = p^*_{\text{i}} $.

In the present work, we essentially assume that the matrix element $ |\mathcal{M}|_\Sigma^2  = |\langle\bm{p}_3\bm{p}_4|\mathcal{M}|\bm{p}_1 \bm{p}_2 \rangle_\Sigma|^2$
does not depend much on the presence of the other particles near the colliding two
particles. However, since the invariant matrix element is defined for the plane waves normalized as
\begin{equation}
\langle\bm{p}|\bm{p}'\rangle_\Sigma=2E^*(\bm{p})(2\pi)^{3}\delta^{3}(\bm{p}-\bm{p}')
\end{equation}
depending on the potential $\Sigma$, we choose to relate $|\mathcal{M}|_\Sigma^2$ to the matrix element $|\mathcal{M}|^2_{\Sigma=0}$ in the vacuum by
\begin{equation}
 |\mathcal{M}|^2_{\Sigma} = 
\biggl[
\frac{E_1^* E_2^*}{\tilde{\omega}_1 \tilde{\omega}_2 }
\biggr]_{\text{in}}
\biggl[
\frac{E_3^* E_4^*}{\tilde{\omega}_3 \tilde{\omega}_4 }
\biggr]_{\text{out}}
 |\mathcal{M}|^2_{\Sigma=0}
\end{equation}
with
\begin{equation}
 \tilde{\omega}_i = \sqrt{m^2_i + \bm{p}^{*2}_i }\qquad(i=1,2,3,4).
\end{equation}
The cross section under the potential is finally written as
\begin{equation}
d\sigma = f_{\text{in}} f_{\text{out}}
\frac{|\mathcal{M}|^2_{\Sigma=0}}{16\pi \tilde{s}}
\frac{[p_{\text{f}}^*]_{\text{out}}}{[p_{\text{i}}^*]_{\text{in}}}
\frac{A(m_4)dm_4}{2\pi} \frac{d\Omega_{\text{f}}^*}{4\pi},
\label{eq:dsigma-pot}
\end{equation}
with
\begin{align}
\tilde{s} &= [ \tilde{\omega}_1 + \tilde{\omega}_2  ]_{\text{in}}
[ \tilde{\omega}_3 + \tilde{\omega}_4 ]_{\text{out}}, \\
f_{\text{in}} &= \Bigl[ \frac{1}{\tilde{\omega}_1} + \frac{1}{\tilde{\omega}_2}  \Bigr]_{\text{in}}
\Big{/} \Bigl[ \frac{1}{E^*_1} + \frac{1}{E^*_2}  \Bigr]_{\text{in}}, \\
f_{\text{out}} &= \Bigl[ \frac{1}{\tilde{\omega}_3} + \frac{1}{\tilde{\omega}_4}  \Bigr]_{\text{out}}
\Big{/} \Bigl[ \frac{1}{E^*_3} + \frac{1}{E^*_4}  \Bigr]_{\text{out}}.
\end{align}

For a given initial condition for $\bm{p}_1$ and $\bm{p}_2$, the cross section depends on the potentials ($\Sigma_{i}^s$, $\Sigma_{i}^0$, $\bm{\Sigma}_i$) for the particles in the initial and final states through the phase space factor $f_{\text{in}} f_{\text{out}} [p_{\text{f}}^*]_{\text{out}}/ [p_{\text{i}}^*]_{\text{in}}$, which we calculate precisely at every chance of collisions. In particular, the final momentum is obtained by
\begin{equation}
[p_{\text{f}}^*]_{\text{out}}=
\sqrt{\frac{[s^*_{\text{out}}-(m^*_3+m^*_4)^2][s^*_{\text{out}}-(m^*_3-m^*_4)^2]}
{4s^*_{\text{out}}}},
\label{eq:pfout}
\end{equation}
where
\begin{equation}
s^*_{\text{out}}=(E^*_3+E^*_4)^2-(\bm{p}^*_3+\bm{p}^*_4)^2
\end{equation}
is determined by the energy and momentum conservation
\begin{align}
E^*_3+E^*_4 &= E^*_1+E^*_2+\Sigma^0_1+\Sigma^0_2-\Sigma^0_3-\Sigma^0_4,\\
\bm{p}^*_3+\bm{p}^*_4 &= \bm{p}_1+\bm{p}_2-\bm{\Sigma}_3-\bm{\Sigma}_4.
\end{align}
The condition $[p^*_{\text{f}}]_{\text{out}}=0$ determines the threshold. When the final state includes a resonance, the threshold can be defined as a function of the resonance mass $m_4$.

\subsection{The $NN \to N \Delta$ process}
For the $NN \to N \Delta$ process in free space, we assume isotropic scattering and use a parametrization of the matrix element 
\begin{equation}
\frac{|\mathcal{M}|^2_{\Sigma=0}}{16\pi s} = B\frac{\Gamma_\Delta^2}{(s-M_\Delta^2)^2+s\Gamma_\Delta^2},
\end{equation}
which is the same form as adopted in the UrQMD model \cite{bass1998} but we take $ B =64400$ mb GeV$^2$, $\Gamma_\Delta=0.118$ GeV and $M_\Delta=1.232$ GeV \cite{nara1999}.  
The dependence on $s$ is moderate in the region of our interest including near the threshold. Therefore, we can allow some arbitrariness in $s$ at which the matrix element should be evaluated when the potentials are present. Considering that the matrix element is essentially a function of the momenta rather than the energies, we choose 
\begin{equation}
s = \tilde{s}_{NN}=[\tilde{\omega}_1+\tilde{\omega}_2]_{\text{in}}^2=4(m_N^2+p_N^{*2})
\label{eq:tildesNN}
\end{equation}
with $p^{*}_N = [p^{*2}_\text{i}]_\text{in}$ being the kinetic momentum of a nucleon in the rest frame of the $NN$ system. Then, from Eq.~\eqref{eq:dsigma-pot}, we write the cross section as
\begin{align}
\sigma_{NN \rightarrow N\Delta} & = C_{NN  N\Delta}
f_{\text{in}} f_{\text{out}}
\biggl(\frac{|\mathcal{M}|^2_{\Sigma=0}}{16\pi s}\biggr)_{\!s=\tilde{s}_{NN}} \nonumber\\
&\quad \times  \frac{[p_{\text{f}}^*]_{\text{out}}}{[p_{\text{i}}^*]_{\text{in}}} 
 \frac{A_\Delta(m)dm}{2\pi}, 
\label{eq_sigma}
\end{align}
where $m$ is the vacuum mass of $\Delta$ and $C_{NN N\Delta}$ is the isospin Clebsh-Gordan factor
\begin{equation}
C_{NNN\Delta}=\begin{cases}
\frac34 & \mbox{for $nnp\Delta^-$, $ppn\Delta^{++}$}\\
\frac14 & \mbox{for $nnn\Delta^0$, $ppp\Delta^+$, $npn\Delta^+$ $npp\Delta^0$}\\
0 & \mbox{otherwise}.
\end{cases}
\end{equation}
The spectral function of the $\Delta$ resonance is parametrized as
\begin{equation}
A_\Delta(m) = \frac{ 4 m^2 \Gamma^{\text{tot}}_\Delta(m)}{(m^2 - M_\Delta^2)^2+ m^2\Gamma^{\text{tot}}_\Delta(m)^2},
\label{eq:Asp}
\end{equation}
where the total width $\Gamma^{\text{tot}}_\Delta(m)$ is determined below in Sec.~\ref{sec:DNpi}, depending again on the potentials in the initial and final states of the $\Delta\to N\pi$ process.

Under the presence of potentials, the $\Delta$ decay width in principle depends on the momentum of $\Delta$, which is determined by the scattering angle in the $NN\to N\Delta$ process. In the present study, we ignore this dependence by using $\Gamma^{\text{tot}}_\Delta(m)$ evaluated for $\Delta$ at rest in the `out' frame.

When there are several reaction channels starting with the same initial channel, e.g., $p+n\to p+\Delta^0$ and $p+n\to n+\Delta^+$, the threshold is channel dependent in general because the potentials in the final particles depend on the channel. Furthermore, the width $\Gamma^{\text{tot}}_\Delta(m)$ and therefore the spectral function $A_\Delta(m)$ also depend on the channel, e.g., whether $\Delta=\Delta^0$ or $\Delta^+$. We will correctly treat such cases, while such channel dependence is often treated approximately in other transport models, e.g., Ref.~\cite{buss2012}.

\subsection{The $N\Delta \to NN$ process}
The inverse process $N\Delta \to NN$ is described by the matrix element that is related to that of the $NN \to N\Delta$ process by
\begin{equation}
 g_{NN} |\mathcal{M}_{NN \to N\Delta}|^2 = g_{N\Delta} |\mathcal{M}_{N\Delta \to NN}|^2,
\end{equation}
where $ g_{NN}$ and $ g_{N\Delta}$ are the spin degeneracy factors, $ g_{NN} = 4$ and $ g_{N\Delta}=8$.
Therefore, we have
\begin{multline}
\sigma_{N\Delta \rightarrow NN} = \frac{g_{NN}}{g_{N\Delta}}
\frac{C_{NN N\Delta}}{1 + \delta_{NN}}
f_{\text{in}} f_{\text{out}}
\biggl(\frac{|\mathcal{M}|^2_{\Sigma=0}}{16\pi s}\biggr)_{\!s=\tilde{s}_{NN}}
\frac{[p_{\text{f}}^*]_{\text{out}}}{[p_{\text{i}}^*]_{\text{in}}},
\end{multline}
in which the factor $1/(1 + \delta_{NN})$ takes into account the limitation in angle integral for a final state with identical particles.

\subsection{The $\Delta \to N\pi$ process\label{sec:DNpi}}
For a decay process of a particle to two particles $1 \to 3+4$, the decay rate in the rest frame of the decaying particle is
\begin{equation}
 \Gamma = \frac{|\mathcal{M}|^2_{\Sigma}}{\pi} \frac{1}{2 m^*_1}
\biggl[ \frac{p^{*2}_\text{f}}{4 v_\text{f} E_3^*E_4^*} \biggr]_{\text{out}}.
\end{equation}
We relate the matrix element in the presence of potentials $\Sigma$ to that in the free space by
\begin{equation}
|\mathcal{M}|_\Sigma^2
=\frac{m_1^*}{m_1}
\biggl[ \frac{E_3^*E_4^*}{\tilde{\omega}_3\tilde{\omega}_4} \biggr]_{\text{out}}
|\mathcal{M}|^2_{\Sigma=0},
\end{equation}
so that
\begin{equation}
 \Gamma = f_{\text{out}} \frac{|\mathcal{M}|^2_{\Sigma=0}}{8 \pi \tilde{s}} [  p^*_\text{f}  ]_{\text{out}}
\end{equation}
with
\begin{equation}
\tilde{s} = m_1 [ \tilde{\omega}_3 +  \tilde{\omega}_4]_\text{out}.
\end{equation}

For the $\Delta \to N \pi$ process, several parametrizations were studied by Weil~\cite{Weil:2013mya}, among which the present work uses the form by Manley et al.~\cite{Manley:1992yb} that corresponds to a choice of the matrix element for the $p$-wave decay as
\begin{equation}
\frac{|\mathcal{M}|^2_{\Sigma=0}}{8\pi\tilde{s}} = 
\frac{M_0 \Gamma_0}{m_\Delta p_0}
\left(\frac{[p^*_\text{f}]_{\text{out}}}{p_0}\right)^2
\frac{ p_0^{2} + \Lambda^2  }{ [p^*_\text{f}]_{\text{out}}^2+  \Lambda^2 },
\end{equation}
where $m_\Delta$ is the vacuum mass of the decaying $\Delta$ and the constant parameters are $M_0=1.232$ GeV, $\Gamma_0 =0.118$ GeV, $\Lambda = 1\ \text{fm}^{-1}$, and
\begin{equation}
p_0=\sqrt{[M_0^2-(m_N+m_\pi)^2][M_0^2-(m_N-m_\pi)^2]/4M_0^2}.
\end{equation}
Considering the isospin Clebsh-Gordan factor, the $\Delta$ decay width in the presence of potentials is
\begin{equation}
 \Gamma_{\Delta \to N \pi}(m_\Delta) = C_{\Delta N \pi} f_{\text{out}} \frac{M_0 \Gamma_0}{m_\Delta} 
\left(\frac{[p^*_\text{f}]_{\text{out}}}{p_0}\right)^3
\frac{ p_0^{2} + \Lambda^2  }{ [p^*_\text{f}]_{\text{out}}^2+  \Lambda^2 },
\end{equation}
with
\begin{equation}
C_{\Delta N\pi}=
\begin{cases}
1 & \mbox{for $\Delta^-\leftrightarrow n\pi^-$, $\Delta^{++}\leftrightarrow p\pi^+$}\\
\frac23 & \mbox{for $\Delta^0\leftrightarrow n\pi^0$, $\Delta^+\leftrightarrow p\pi^0$}\\
\frac13 & \mbox{for $\Delta^0\leftrightarrow p\pi^-$, $\Delta^+\leftrightarrow n\pi^+$}\\
0 & \mbox{otherwise}.
\end{cases}
\end{equation}
Starting with the same $\Delta$ mass $m_\Delta$, the final momentum $[p^*_{\text{f}}]_{\text{out}}$ and therefore the decay width depend on the decay channel, e.g., whether $\Delta^0\to n+\pi^0$ or $\Delta^0\to p+\pi^-$, not only due to the Clebsh-Gordan factor but also because of the different potentials of particles in different final channels.

As for the total width, in the spectral function $A_\Delta(m)$ of Eq.~\eqref{eq:Asp}, we use
\begin{equation}
\label{eq:TotalGam_Delta}
 \Gamma^{\text{tot}}_\Delta(m) = \Gamma_{\text{sp}}^\Delta\frac{\rho}{\rho_0} + \sum \Gamma_{\Delta \to N \pi}(m),
\end{equation} 
where the second term in the r.h.s.\ is the $\Delta$ decay width summed over the isospin channels of $N + \pi$. The first term in the r.h.s.\ of Eq.~\eqref{eq:TotalGam_Delta} represents the in-medium $\Delta$ spreading width due to the absorption and rescattering processes such as $\Delta N \to NN$ and $\Delta N \to \Delta N$, as considered by Larionov et al.~\cite{larionov2003}.
In the present work, we take the parameter $\Gamma_{\text{sp}}^\Delta = 60$ MeV as a default setting.

\subsection{The $N \pi \to \Delta$ process}
The cross section for $3 + 4 \to 1$ is
\begin{equation}
d\sigma =  \biggl[ \frac{1}{4 v_{\text{i}} E^*_3 E^*_4} \biggr]_\text{in}
|\mathcal{M}|^2_{\Sigma} \frac{A(m_1)dm_1}{2\pi} 
\frac{2\pi\delta(E_1 - E_3 - E_4)}{2E^*_1}.
\end{equation}
By integrating this over $m_1$, we have
\begin{equation}
\sigma = f_{\text{in}}
\frac{|\mathcal{M}|^2_{\Sigma=0}}{8\pi\tilde{s}}
\frac{\pi}{[p_{\text{i}}^*]_{\text{in}}}  A(m_1),
\end{equation}
where $m_1$ is determined by the energy conservation.

For the $N \pi \to \Delta$ process, the matrix element is related to that of the inverse process by
\begin{equation}
 g_{N\pi} |\mathcal{M}_{N\pi \to \Delta}|^2 = g_{\Delta} |\mathcal{M}_{\Delta \to N\pi}|^2
\end{equation}
with the spin degeneracy factors $ g_{N\pi} = 2$ and $ g_{\Delta}=4$, and therefore the cross section is related to the decay rate as
\begin{equation}
 \sigma_{N\pi \rightarrow \Delta} = \frac{g_{\Delta}}{g_{N\pi}}
\frac{\pi}{[p_{\text{i}}^*]^2_{\text{in}}}
\Gamma_{\Delta \to N \pi}(m_\Delta) A(m_\Delta).
\end{equation}

\section{The $NN \rightarrow N \Delta$ cross sections in asymmetric nuclear matter \label{sec:nnnd}}

\begin{figure*}
\centering
\includegraphics[scale=0.32]{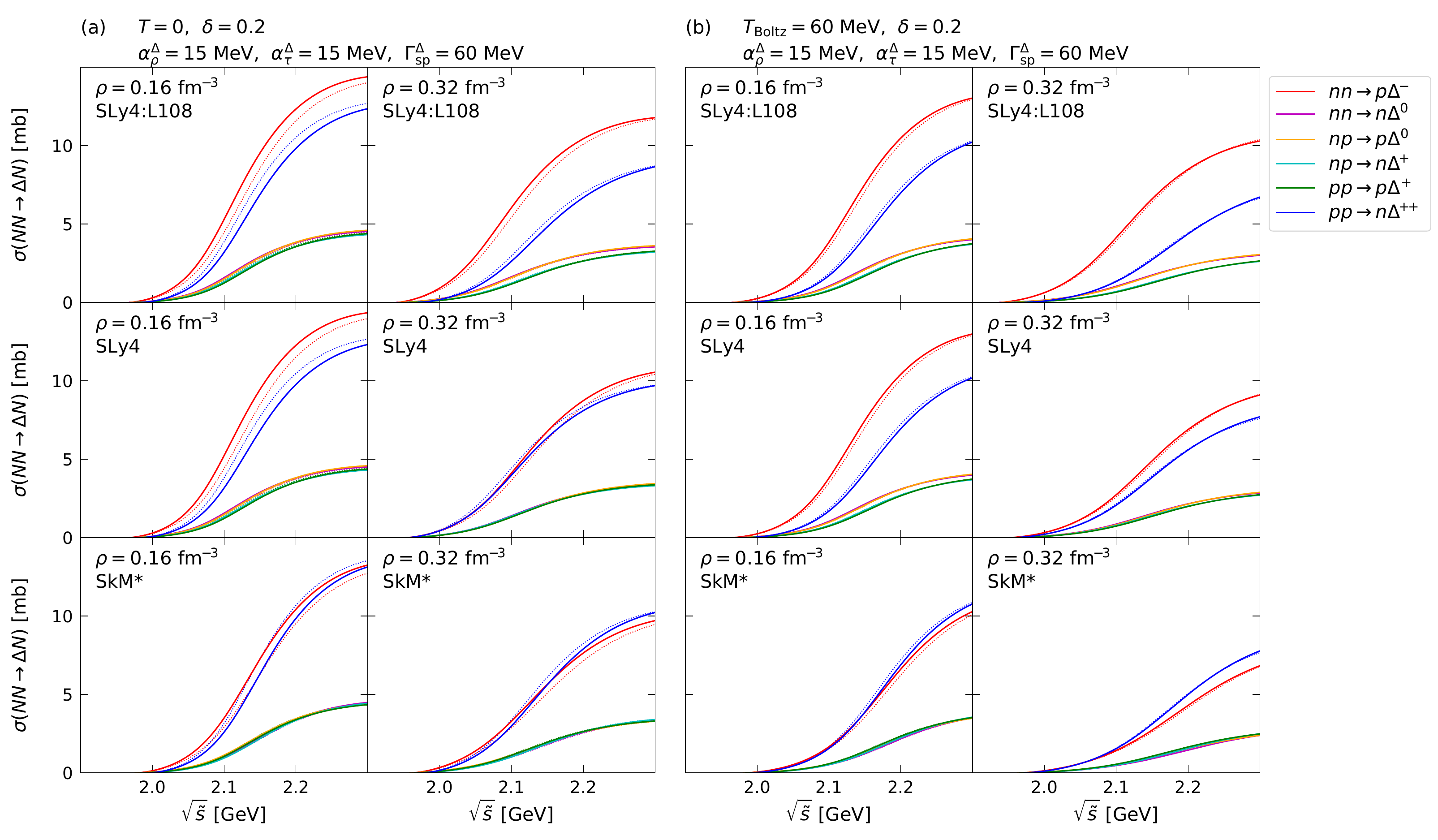}
\caption{ Left part (a): The $NN \rightarrow N \Delta$ cross sections for different channels as a function of $\sqrt{\tilde{s}}$ (see text) in the asymmetric nuclear matter $\delta = 0.20$ at the zero temperature $T=0$, for the three cases of the nuclear interaction based on SLy4:L108 (top), SLy4 (middle) and SkM* (bottom). The left column is for the density $\rho=\rho_0$ and the right column is for $\rho=2\rho_0$. The two cases of the isovector part of the $\Delta$ potential are shown with the solid lines ($\gamma^\Delta=1$) and the dotted lines ($\gamma^\Delta =3$). 
The isoscalar part of the $\Delta$ potential includes repulsive terms with the $\alpha_\rho^\Delta = 15$~MeV, $\alpha_\tau^\Delta = 15$~MeV, and the spreading width parameter is $\Gamma_{\rm sp}^\Delta= 60$~MeV.
Right part (b): The same as (a), but at a finite temperature corresponding to the kinetic energy density $\tau_b=3m_NT_{\text{Boltz}}\rho_b$ with $T_{\text{Boltz}}= 60$~MeV. 
}
\label{fig:cross}
 \end{figure*}

In this section, we give discussions on some examples of the cross sections in the nuclear matter to understand the features of the $NN\to N\Delta$ cross sections under the presence of potentials.
 The cross sections shown here are calculated in the same formalism as in the heavy-ion collision simulations in Sec. \ref{sec:AMDsJAM}.

In the left part (a) of Fig.~\ref{fig:cross}, we show the $NN \rightarrow N \Delta$ cross sections for different channels of the $\Delta$ production under the presence of potentials. The initial two nucleons with momenta $\pm\bm{p}_N$ are placed in the nuclear matter with the isospin asymmetry $\delta = 0.20$ and the temperature $T=0$. The nucleon potentials are chosen as described in Sec.~\ref{sec:Potamd} for the three cases based on the Skyrme parametrizations SLy4:L108 (top), SLy4 (middle) and SkM* (bottom). As for the $\Delta$ potential of Eq.~(\ref{eq:Sigma_Delta}), this figure shows the case when the isoscalar part is taken as $\Sigma_{\text{is}}=\frac12(\Sigma_n+\Sigma_p)_{\text{SkM*}}$ with additional repulsive terms ($\alpha_\rho^\Delta = 15$ MeV and $\alpha_\tau^\Delta = 15$ MeV). 
For the isovector part $\Sigma_{\rm iv}$, two cases of the isospin splitting parameter are shown for $\gamma^\Delta =1$ (solid lines) and $\gamma^\Delta=3$ (thin dotted lines). The spreading width of $\Delta$ is taken into account with $\Gamma_{\text{sp}}^\Delta=60$ MeV in Eq.~\eqref{eq:TotalGam_Delta}. Note that the cross sections are shown here as functions of $\sqrt{\tilde{s}}=\tilde{s}_{NN}^{1/2}=2(m_N^2+p_N^2)^{1/2}$ which is a direct function of $p_N=|\bm{p}_N|$ without dependence on potentials.

The effect of the isospin asymmetry ($\delta=0.2$) is evident in the cross sections of different isospin channels of $NN\to N\Delta$ in Fig.~\ref{fig:cross} (a), in particular in the difference between $nn\to p\Delta^-$ (red) and $pp\to n\Delta^{++}$ (blue). This channel dependence is relatively small in the SkM* case of the nucleon potentials compared to the SLy4 and SLy4:L108 cases at the density $\rho=0.16\ \text{fm}^{-3}$ in the left column. When the density is raised to $\rho=0.32\ \text{fm}^{-3}$ in the right column, the channel dependence is particularly large in the SLy4:L108 case, and the weak channel dependence in SkM* is further weakened or even inverted.
The same behaviors are observed under a high temperature condition as shown in the right part (b) of Fig.~\ref{fig:cross}.

When the cross sections in different channels are compared at the same $\sqrt{\tilde{s}}$, i.e., at the same initial nucleon momentum $p_N$, the channel dependence may be understood based on the most important factor \begin{equation}
\epsilon^*\equiv(s^*_{\text{out}})^{1/2}-(m_3^*+m_4^*)
\end{equation}
in Eq.~\eqref{eq:pfout} for $[p^*_{\text{f}}]_{\text{out}}$, which we can write in the present case of nuclear matter as
\begin{align}
\epsilon^*&=\sqrt{m_1^{*2}+p_N^2}+\sqrt{m_2^{*2}+p_N^2} \nonumber\\
&\quad +\Sigma_1^0+\Sigma_2^0-\Sigma_3^0-\Sigma_4^0-m_3^*-m_4^* \nonumber\\
&= \epsilon_{\text{free}}(m_4)+\Delta U(p_N)
\end{align}
with
\begin{align}
\epsilon_{\text{free}} (m_4)&=2\sqrt{m_N^2+p_N^2}-m_N-m_4,\\
\Delta U(p_N) &= U_1(p_N)+U_2(p_N)-U_3(0)-U_4(0).\label{eq:DU}
\end{align}
Here, $U_i(p)$ ($i=1,2,3$) are the momentum dependent nucleon potentials defined by Eq.~\eqref{eq:Urel} and $U_4(0)$ is that of $\Delta$ at zero momentum. It should be noted that the initial nucleon momentum $p_N$, at which $U_1$ and $U_2$ are evaluated in $\Delta U(p_N)$, has to be high ($p_N\gtrsim 500$ MeV/$c$) for a production of $\Delta$, while the potential $U_3$ of the final nucleon is evaluated at $p=0$. When we can ignore the isospin splitting in $U_4$ for $\Delta$, we can understand the channel dependence of cross sections in Fig.~\ref{fig:cross} based on the difference between $U_1(p_N)+U_2(p_N)$ and $U_3(0)$, which we can read from Fig.~\ref{fig:pot}. For example, let us compare the $nn\to p\Delta^-$ and $pp\to n\Delta^{++}$ processes. In the case of SLy4:L108, Fig.~\ref{fig:pot} shows $U_n(p)>U_p(p)$ at all momenta and therefore $2U_n(p_N)-U_p(0)$ for $nn\to p\Delta^-$ is greater than $2U_p(p_N)-U_n(0)$ for $pp\to n\Delta^{++}$, which can explain the strong channel dependence in Fig.~\ref{fig:cross}. The SLy4 case is identical to the SLy4:L108 case at $\rho=0.16\ \text{fm}^{-3}$, while at a higher density $\rho=0.32\ \text{fm}^{-3}$ the neutron and proton potentials are similar $U_n(p)\approx U_p(p)$, which makes the channel dependence weak in the cross sections at the high density. In the case of SkM*, the channel dependence of the cross sections is relatively weak or even inverted compared to the SLy4 case because $U_n(p_N)\approx U_p(p_N)$ or $U_n(p_N)<U_p(p_N)$ at high momenta while $U_n(0)>U_p(0)$ at zero momentum. 

We can appreciate the effect of the isovector part $\Sigma_{\text{iv}}$ of the $\Delta$ potential by comparing the $\gamma^\Delta=3$ case of the isospin splitting parameter (thin dotted line) to the $\gamma^\Delta=0$ case (solid line) in each channel. In the case of Fig.~\ref{fig:cross} (a) at zero temperature, a strong splitting ($\gamma^\Delta=3$) results in a weakening of the channel dependence under most of these asymmetric conditions. We may understand this because the splitting now enters in $U_4(0)$ in $\Delta U(p_N)$ of Eq.~\eqref{eq:DU}. In contrast, in the case of Fig.~\ref{fig:cross} (b) at a high temperature, we find that the isospin splitting of the $\Delta$ potential influences the cross sections only weakly. This may be partly because the isospin splitting between $U_\Delta$ at the high temperature (Fig.~\ref{fig:pot} (b)) is smaller than that at zero temperature (Fig.~\ref{fig:pot} (a)) when compared at the same splitting factor $\gamma^\Delta$.

The condition $\epsilon^*=0$ determines the threshold for the production of $\Delta$ at a vacuum mass $m_4$, and thus the threshold momentum $p_{N,\text{th}}(m_4)$ or $\tilde{s}_{\text{th}}(m_4)$ in each channel can be defined as a function of $m_4$. This threshold naturally depends on the channel through the isospin dependence of the potentials of nucleons and $\Delta$ in $\Delta U(p_N)$. To the authors' knowledge, this kind of threshold effect was argued by Fermini et al.~\cite{ferini2005} and by Cozma~\cite{cozma2016}. On the other hand, the minimum value of $p_N$ or $\tilde{s}$ for the $\Delta$ production that can be read for each channel from Fig.~\ref{fig:cross} is $p_{N,\text{th}}(m_{\text{min}})$ or $\tilde{s}_{\text{th}}(m_{\text{min}})$, which is the threshold to produce $\Delta$ at the minimum mass $m_{\text{min}}$. In our framework, the minimum mass is determined by the condition $[p_{\text{f}}^*]_{\text{out}}=0$ for the decay process $\Delta\to N+\pi$, i.e.,
\begin{equation}
m_{\text{min}}+U_\Delta(0)=m_N+U_N(0)+m_\pi,
\end{equation}
where $U_N(0)$ is for the nucleon after the decay of $\Delta$.  Using this relation, the threshold condition $\epsilon^*=0$ with $m_4=m_{\text{min}}$ and $U_4(0)=U_\Delta(0)$ is now obtained as
\begin{multline}
2\sqrt{m_N^2+[p_{N,\text{th}}(m_{\text{min}})]^2}-2m_N-m_\pi \\
 + U_1(p_N)+U_2(p_N)-U_3(0)-U_N(0)=0.
\end{multline}
Therefore, the threshold, $p_{N,\text{th}}(m_{\text{min}})$ or $\tilde{s}_{\text{th}}(m_{\text{min}})$, depends on nucleon potentials but does not depend on the choice of the $\Delta$ potential. This kind of threshold was considered by the authors of Refs.~\cite{zhenzhang2017,cui2018}. In our case, by a careful look at Fig.~\ref{fig:cross}, we can confirm that the threshold of each channel does not depend on the choice of the parameter $\gamma^\Delta$ for the isospin splitting of the $\Delta$ potential. This also implies that the shift of the threshold, such as an assumption like $\sigma(\sqrt{\tilde{s}})=\sigma_{\text{free}}(\sqrt{\tilde{s}}-\text{const.})$, is not sufficient to express the effects of potentials in the cross sections, as also can be found in the results of Refs.~\cite{cui2018,larionov2003} with the one-boson exchange model at the zero temperature for the symmetric nuclear matter~\cite{larionov2003} and the asymmetric nuclear matter~\cite{cui2018}.

\section{Simulation of heavy-ion collisions in the AMD+sJAM model\label{sec:AMDsJAM}}
In the present work, we first solve the dynamics of neutrons and protons by antisymmetrized molecular dynamics (AMD)~\cite{ikeno2016,ono2013nn,ono2019review}.
AMD describes the dynamics of a many-nucleon system by the time evolution of a Slater determinant of Gaussian wave packets.
We use the nuclear effective energy density functional given by Eq.~\eqref{eq:Edensity} in Sec.~\ref{sec:Potamd}.
The AMD model has some advantages in that antisymmetrization is treated accurately and cluster correlations can be taken into account by extending the two-nucleon collision process~\cite{ikeno2016,ono2013nn}.
When two nucleons $N_1$ and $N_2$ collide, we consider the process
\begin{equation}
N_1+N_2+B_1+B_2\rightarrow C_1+C_2 \,,
\end{equation}
where each of the scattered nucleons $N_j$ ($j=1,2$) may form a cluster $C_j$ (up to $\alpha$ cluster) with a spectator particle $B_j$ (nucleon or cluster). This includes the special cases in which both or one of $B_1$ and $B_2$ is empty, e.g., $N_1+N_2\to N_1+N_2$ and $N_1+N_2+B_1\to C_1+N_2$.
The cross section of the process to form clusters $(C_1,C_2)$ is given by
\begin{equation}
\frac{d\sigma(C_1,C_2)}{d\Omega}=
P(C_1,C_2,p_{\text{f}},\Omega)
\frac{p_{\text{i}}}{v_{\text{i}}}
\frac{p_{\text{f}}}{v_{\text{f}}}
|M|^2
\frac{p_{\text{f}}}{p_{\text{i}}},
\end{equation}
where $p_{\text{i}}$ and $v_{\text{i}}$ are the initial relative momentum and the velocity between the colliding nucleons $N_1$ and $N_2$. The relative momentum vector after the momentum transfer between them is denoted by $(p_{\text{f}},\Omega)$, and $p_{\text{f}}$ is determined to conserve the energy $E$ of the system which includes the adopted effective interaction. The velocity factor $v_{\text{f}}=\partial E/\partial p_{\text{f}}$ as a function of $p_{\text{f}}$ also depends on the effective interaction. 
Here, $E$ is the total energy of the system obtained for the antisymmetrized wave function in the AMD calculation. The values of $p_\text{f}$ and $v_\text{f}$ that conserve the energy are determined numerically after several steps of iteration.
The overlap probability factor for cluster formation, $P(C_1,C_2,p_{\text{f}},\Omega)$, is defined by considering the non-orthogonality between the states of different configurations~(Refs.~\cite{ono2013nn} and \cite{ikeno2016}).
The matrix element $|M|^2$ for the two-nucleon scattering is directly related to the assumed in-medium cross sections $\sigma_{NN}$.  We may express it as $|M|^2=(2/m_N)^2 d\sigma_{NN}/d\Omega$ where the right-hand side is evaluated at an average of $p_{\text{i}}$ and $p_{\text{f}}$.
The observables of light fragments in the S$\pi$RIT experimental data have been analyzed by AMD calculations~\cite{SpRIT:2021dvt,Lee:2022,NishimuraSpRIT}. 
In the present work, we use the same nucleon calculations as in the analysis of Ref.~\cite{NishimuraSpRIT}.

In the AMD model, however, $\Delta$ resonances and pions have not been incorporated. 
Considering the small pion multiplicity in heavy-ion collisions of our interest, we can still use the nucleon dynamics calculated by AMD, by regarding $\Delta$ and pion production as perturbation. In the AMD+JAM model~\cite{ikeno2016,ikeno2016erratum,Ikeno:2019mne} by Ikeno, Ono, Nara, and Ohnishi, the nucleon dynamics was solved by AMD and then reactions related to pions and $\Delta$ resonances were handled by a hadronic cascade model (JAM). 
 The JAM model is a reliable hadron transport model developed by Nara, Otuka, Ohnishi, Niita, and Chiba~\cite{nara1999}. 
The cascade method by JAM has a feature that the sequence of the collision and decay processes in a many-particle system is handled precisely in the order of the time at which each process should take place. This feature is advantageous in avoiding unphysical dependence on the computational time step parameter, as we found in a comparison of transport models in Ref.~\cite{ono2019} for pion production in a box.
However, in the AMD+JAM calculations, the potentials were not taken into account in the processes related to $\Delta$ resonances and pions.

For a precise treatment of potentials in the collision term, we have developed a new transport code sJAM. This code precisely follows the cascade mode of the JAM code~\cite{nara1999} in the case of the presence of only nucleons, $\Delta$ resonances and pions without any effect of potentials. 
The sJAM code now takes into account the potentials that affect the cross sections and decay rates as formulated in Sec.~\ref{sec:crosssection}. 
The final momenta of a collision or decay in the simulation are determined from Eq.~\eqref{eq:pfout} to conserve the energy.
The potentials also affect the propagation in sJAM through the dispersion relation. However, the 
nuclear force acting on a $\Delta$ resonance is ignored assuming that the momentum change is small during a short time between a production of $\Delta$ and its decay or absorption. 

The electromagnetic force acting on charged particles, including pions, is taken into account in sJAM.
The Lorentz force acting on a charged particle labeled by $i$ is calculated as 
\begin{equation}
 \frac{d p_i^\mu}{d t} = e Z_i F_i^{\mu \nu} \frac{d x_{i, \nu}}{d t}
,\qquad
F_i^{\mu\nu}=\sum_{j(\ne i)}F_{ij}^{\mu\nu},
\end{equation}
by using the electromagnetic field tensor $F_i^{\mu\nu}$ at the space-time point of the particle $x_{i,\nu}=(t,-x_i,-y_i,-z_i)$. 
The field $F_{ij}^{\mu\nu}$ created by the $j$th particle is obtained by first assuming the electrostatic Coulomb field $\bm{E}'= e Z_j\bm{r}_{ij}'/(r_{ij}')^3$ and $\bm{B}'=0$ in the rest frame of the $j$th particle and then Lorentz transforming the field tensor to the reference frame. However, when the distance $r_{ij}'=|\bm{r}_{ij}'|$ in the rest frame of the $j$th particle is less than 2~fm, the field is replaced by zero.
In the AMD code, the Coulomb interaction is considered by solving the Poisson equation.

In the practical AMD+sJAM calculations, the information on the nucleon dynamics calculated by AMD is sent to the sJAM calculation in the form of a list of test particles $(\bm{r}_1,\bm{p}_1)$, $(\bm{r}_2,\bm{p}_2)$, \dots, $(\bm{r}_A,\bm{p}_A)$ at every time step of 1 fm/$c$. These test particles are generated randomly by the method of Ref.~\cite{ikeno2016} according to the Wigner distribution function $f_\alpha(\bm{r},\bm{p})$ in the AMD calculation.
In addition, to allow the calculation with potentials in sJAM, the information on the densities at the positions of test particles, $\rho_n(\bm{r}_i)$, $\rho_p(\bm{r}_i)$, $\bm{J}_n(\bm{r}_i)$, $\bm{J}_p(\bm{r}_i)$, $\tau_n(\bm{r}_i)$ and $\tau_p(\bm{r}_i)$ for $i=1,2,\ldots, A$, are sent from AMD to sJAM. Using these densities, the potentials $\Sigma_a^s(\bm{r}_i)$, $\Sigma_a^0(\bm{r}_i)$ and $\bm{\Sigma}_a(\bm{r}_i)$ are calculated either in AMD or sJAM. Then, as formulated in Sec.~\ref{sec:crosssection}, the cross sections and resonance decay rates are calculated at every chance of collisions and decays. For this purpose, we need to know not only the potential that the particle $i$ currently feels but also the potentials that it may feel when it is changed to other species $a=n,p,\Delta^-,\Delta^0,\Delta^+$ and $\Delta^{++}$ in the final channels of inelastic processes.

Thus, the $NN \leftrightarrow N\Delta$ and $\Delta \leftrightarrow N \pi$ processes are calculated in sJAM under the presence of potentials with a precise treatment of energy conservation. The elastic $NN$ collisions are also considered in sJAM, but the nucleon information is updated at every time step by AMD.

The Pauli blocking factor for the nucleon(s) in the final state of $NN\leftrightarrow N\Delta$ and $\Delta\to N\pi$ processes is determined by using the Wigner function $f(\bm{r}_i,\bm{p}_i')$ calculated precisely for the many-nucleon Slater determinant in AMD. This is the best among the methods investigated in Ref.~\cite{Ikeno:2019mne}. However, in the evaluation of the width parameter $\Gamma^{\text{tot}}_\Delta(m)$ in Eq.~\eqref{eq:Asp} for the spectral function $A_\Delta(m)$, the Pauli blocking in the $\Delta\to N\pi$ final states is ignored.

\begin{figure}
\includegraphics[scale=0.4]{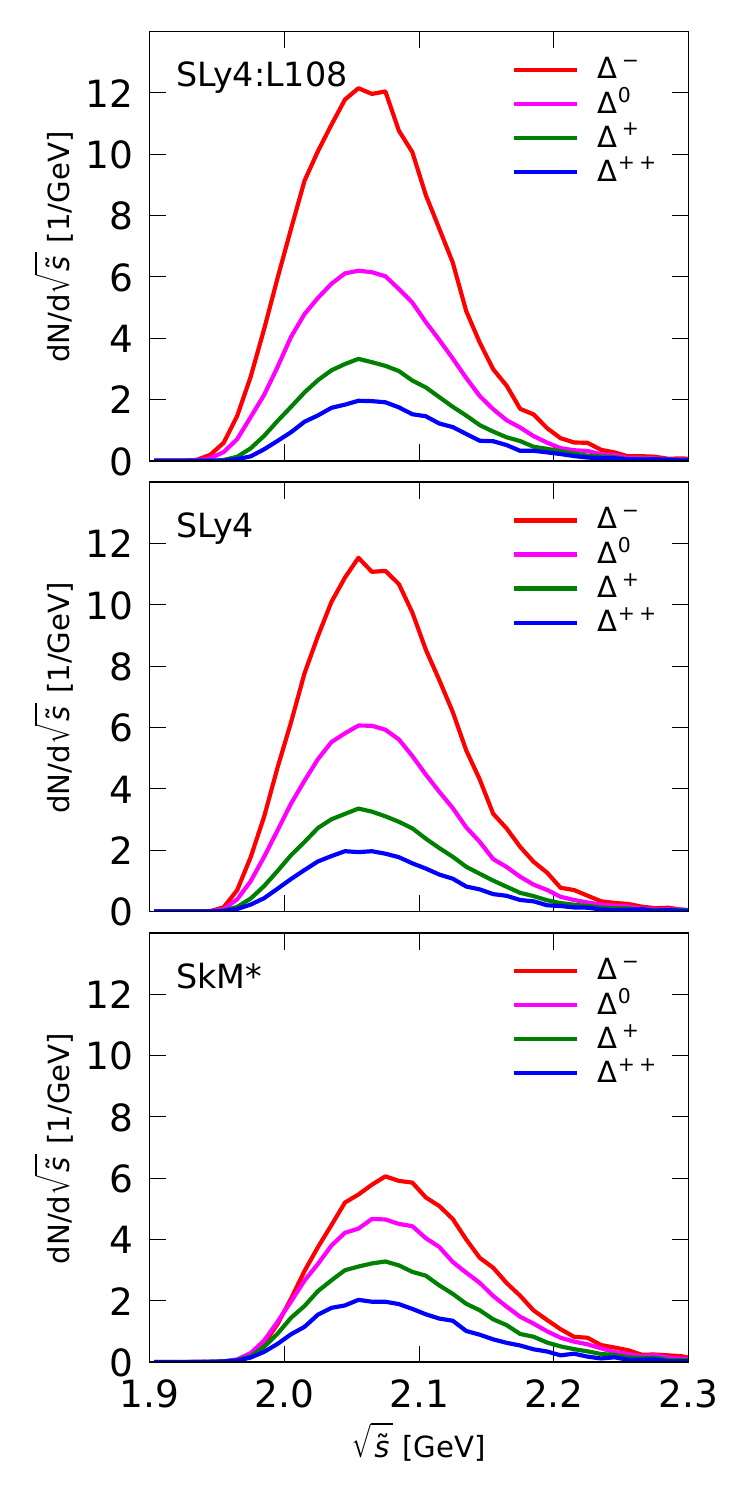}
\caption{ The distribution of $\sqrt{\tilde{s}}$ (see text) for the number of $NN \to N \Delta$ reactions per event in the AMD+sJAM calculations, with the three cases of nucleon interaction based on SLy4:L108 (top), SLy4 (middle), and SkM* (bottom). }
\label{fig:NNND}
\end{figure}

Figure \ref{fig:NNND} shows the $\Delta$ production in the calculation of ${}^{132}\mathrm{Sn}+{}^{124}\mathrm{Sn}$ collisions at $E/A=270$ MeV and for the impact parameter range $b<3$ fm. For the production of $\Delta^-$, $\Delta^0$, $\Delta^+$ and $\Delta^{++}$, the numbers of $NN \to N \Delta$ reactions per event are shown in each panel as a function of $\sqrt{\tilde{s}}=\sqrt{\tilde{s}_{NN}}$ [Eq.~\eqref{eq:tildesNN}]. The three cases of nucleon interaction are shown for the cases based on SLy4:L108 (top panel), SLy4 (middle) and SkM* (bottom). In all cases, the $\Delta$ production peaks between $\sqrt{\tilde{s}}=2.05$ and 2.10 GeV, from which we can appreciate what part in $\sqrt{\tilde{s}}$ is important in the example of $NN\to N\Delta$ cross sections shown in Fig.~\ref{fig:cross}. We can find that the differences in the $\Delta$ production between the three cases of nuclear interaction are well associated with those in the $NN\to N\Delta$ cross sections, in terms of the channel dependence and the absolute value.
By integrating the distribution, we notice that the $\Delta$ production occurs only a few times per event in this heavy-ion collision system, which justifies our perturbative treatment of the AMD+sJAM model.

Due to Eq.~\eqref{eq:pfout}, our calculation conserves the sum of the single-particle energies of the particles involved in a collision or decay, which generally guarantees the total energy conservation by the collision term, thanks to the definition of the single-particle energies by Eq.~\eqref{eq:upot_def}.
However, we have to note that we replaced the original single-particle potential of Eq.~\eqref{eq:upot} by its relativistic form of Eq.~\eqref{eq:Urel} in sJAM, which may affect the conservation of the total energy defined with Eq.~\eqref{eq:Edensity}.
We checked this point by performing a box calculation~\cite{cozma2023} with an interaction similar to that used in the present work, and confirmed that the total energy of the system in the time evolution is conserved within 0.3~MeV per baryon.

\section{Pion observables\label{sec:results}}

\subsection{Pion spectra}

\begin{figure}
\centering
\includegraphics[width=\columnwidth]{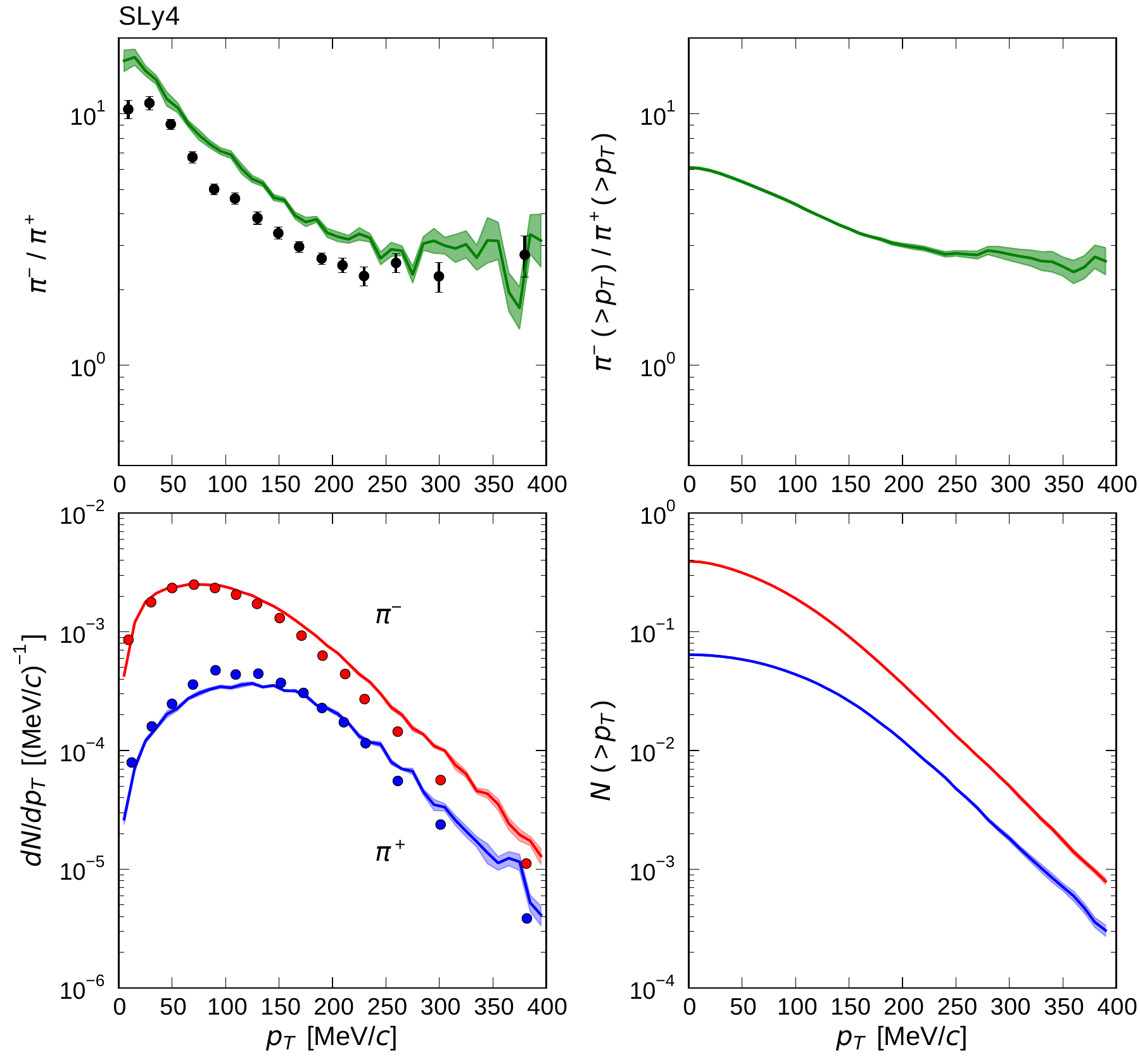}
\caption{Left panels: The transverse momentum ($p_T$) spectra of $\pi^-$ and $\pi^+$ (bottom) and their $\pi^-/\pi^+$ ratio (top), for pions emitted to forward angles $\theta_{\rm c.m.}< 90^\circ$ from the ${}^{132}\mathrm{Sn}+{}^{124}\mathrm{Sn}$ collisions at $E/A=270$~MeV and $b<3$~fm. The calculations are done with the nuclear interaction based on the SLy4 parametrization and with the $\Delta$ potential with  parameters $\alpha_\rho^\Delta = 15$~MeV, $\alpha_\tau^\Delta = 15$~MeV, and $\gamma^\Delta = 1$. The in-medium spreading width of $\Delta$ is taken into account with $\Gamma_{\text{sp}}^\Delta= 60$~MeV.
The experimental data shown by points are taken from Ref.~\cite{spirit2021PRL}.
Right panel: Pion spectra integrated in the momentum range above $p_T$ (bottom, see Eq.~\eqref{eq:integratedspectrum}) and their $\pi^-/\pi^+$ ratio (top).
}
\label{fig:pionspectra-sly4}
\end{figure}

\begin{figure}
\centering
\includegraphics[width=\columnwidth]{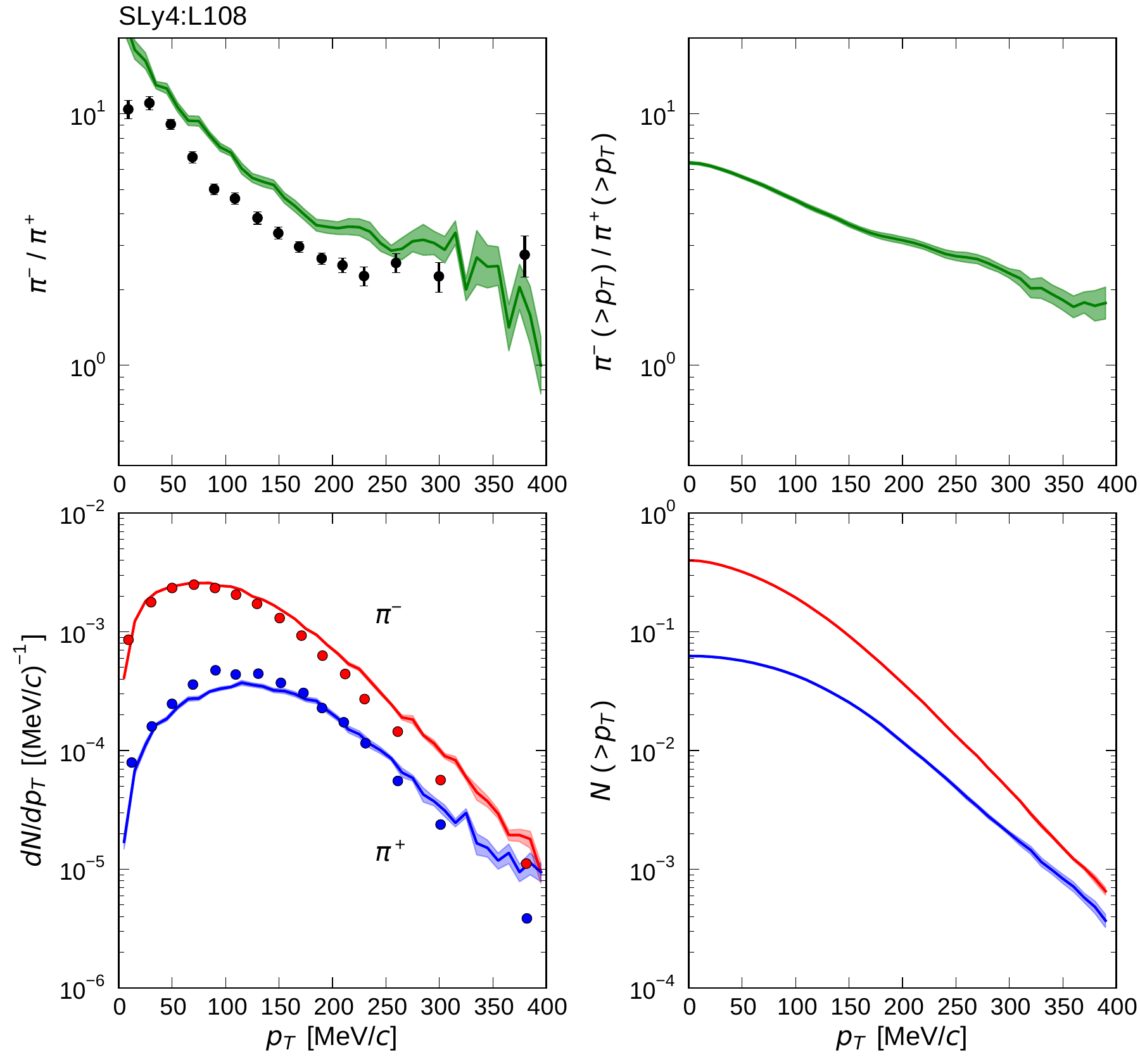}
\caption{The same as Fig.~\ref{fig:pionspectra-sly4} except with the nucleon interaction based on SLy4:L108.
}
\label{fig:pionspectra-l108}
\end{figure}

\begin{figure}
\includegraphics[width=\columnwidth]{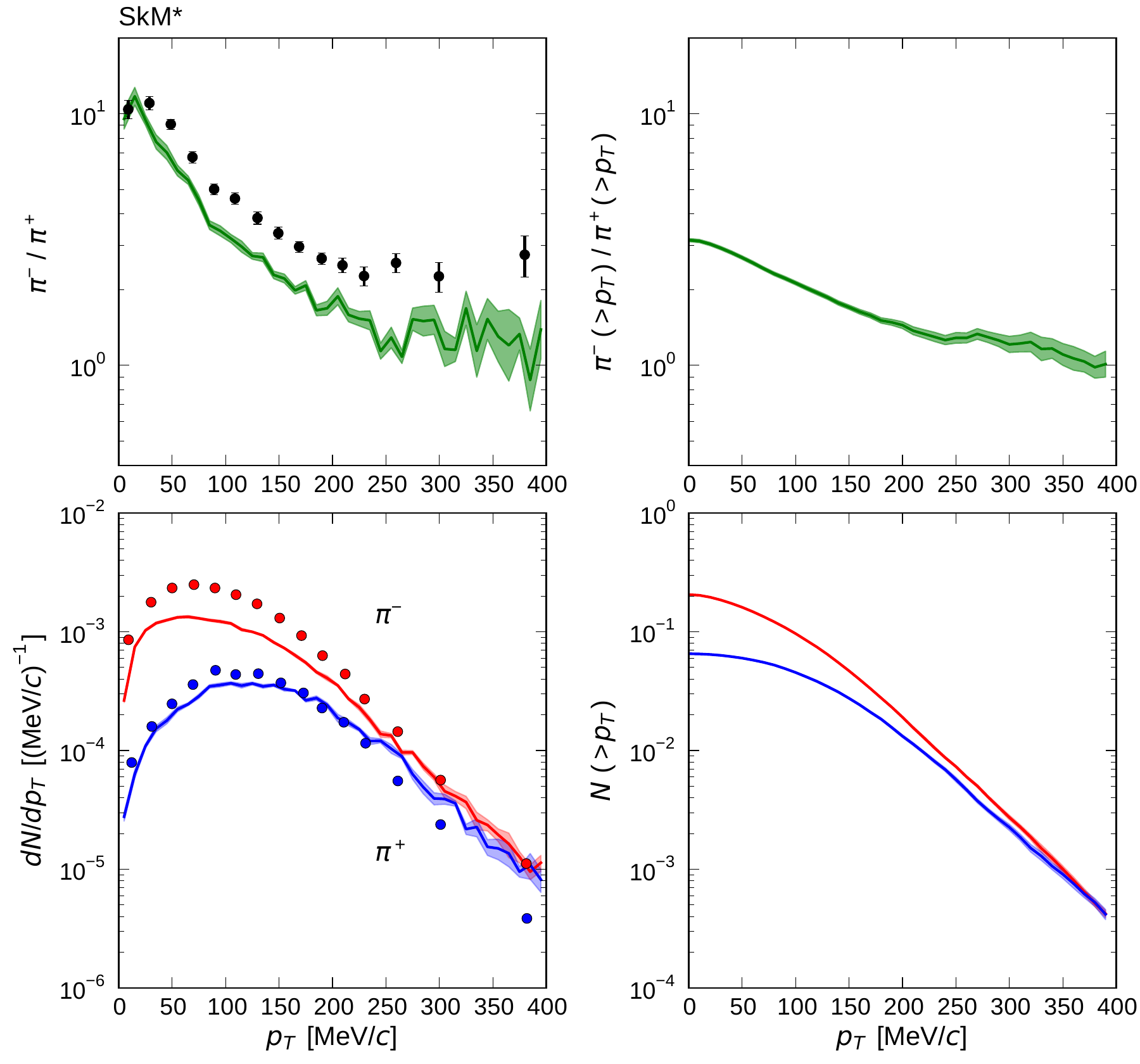}
\caption{The same as Fig.~\ref{fig:pionspectra-sly4} except with the nucleon interaction based on SkM*.
}
\label{fig:pionspectra-skms}
\end{figure}

We calculate the pion production in ${}^{132}\mathrm{Sn}+{}^{124}\mathrm{Sn}$ collision at $E/A=270$~MeV for the impact parameter range $b<3$~fm which corresponds to the S$\pi$RIT experimental data published in Ref.~\cite{spirit2021PRL}.  In Figs.~\ref{fig:pionspectra-sly4}, \ref{fig:pionspectra-l108} and \ref{fig:pionspectra-skms}, we show calculated pion observables in the three cases of nucleon interaction based on SLy4, SLy4:L108 and SkM*, respectively. Note that the momentum dependence is modified by a parameter $\Lambda_{\text{md}}/\hbar=5.0\ \text{fm}^{-1}$ in the AMD calculation and the nucleon potential is converted to a relativistic form in sJAM (see Sec.~\ref{sec:Potamd}). As explained in Sec.~\ref{sec:Dpot}, the $\Delta$ potential, which is parametrized independently of the choice of the nucleon interaction, includes additional repulsive terms with parameters $\alpha_\rho^\Delta = 15$~MeV and $\alpha_\tau^\Delta = 15$~MeV, and an isovector term with the parameter $\gamma^\Delta = 1$.


In these figures, the lines in the bottom left panel show the calculated spectra $dN/dp_T$ of charged pions ($\pi^-$ and $\pi^+$) emitted to forward angles $\theta_{\text{c.m.}}<90^\circ$, as a function of the transverse momentum $p_T$, in comparison with the experimental data of Ref.~\cite{spirit2021PRL} shown by points. The top left panel shows the $\pi^-/\pi^+$ ratio of these spectra. The bottom right panel shows an integral of the spectrum
\begin{equation}
N({>}p_T) = \int_{p_T}^\infty \frac{dN}{dp_T}(p_T')dp_T',
\label{eq:integratedspectrum}
\end{equation}
that is the number of pions emitted with a transverse momentum greater than $p_T$. This quantity is useful to argue the high momentum part of the spectra with a good estimation of the statistical accuracy. The top right column shows the $\pi^-/\pi^+$ ratio of these integrated spectra.

By comparing Fig.~\ref{fig:pionspectra-sly4} (SLy4-based) and Fig.~\ref{fig:pionspectra-l108} (SLy4:L108-based), we can argue the effect of the density dependence of the symmetry energy. However, the difference is not large in the pion yields and spectra between these cases of $L=46$ and 108 MeV. We will discuss this point in detail in the next subsection. The calculated pion results in these cases are similar to the experimental data. In particular, the $\pi^-/\pi^+$ ratio is higher than the experimental data, in contrast to the much smaller $\pi^-/\pi^+$ ratio predicted by AMD+JAM in Refs.~\cite{ikeno2016,ikeno2016erratum,Ikeno:2019mne} and in Ref.~\cite{spirit2021PLB}, where the potentials were not taken into account in $NN\leftrightarrow N\Delta$ and $\Delta\leftrightarrow N\pi$ processes. The large $\pi^-/\pi^+$ ratio in the present calculation can be roughly associated with the strong channel dependence of the $NN\to N\Delta$ cross sections illustrated for nuclear matter in the upper two panels of Fig.~\ref{fig:cross}, which is further related to the behavior of the momentum dependence of the neutron and proton potentials in Fig.~\ref{fig:pot} (see Sec.~\ref{sec:Potamd}).

On the other hand, by comparing Fig.~\ref{fig:pionspectra-skms} (SkM*-based) to Fig.~\ref{fig:pionspectra-sly4} (SLy4-based), we can appreciate the effect of the momentum dependence of the neutron and proton potentials ($U_n$ and $U_p$) under a common condition on the symmetry energy ($L=46$ MeV). As illustrated in Sec.~\ref{sec:nnnd} for nuclear matter, the channel dependence of $NN\to N\Delta$ cross sections is weak or even inverted in the case based on SkM* (bottom panels of Fig.~\ref{fig:cross}) due to the weak momentum dependence of $U_n$ compared to that of $U_p$ (bottom panels of Fig.~\ref{fig:pot}). This can explain the small $\pi^-/\pi^+$ ratio in the case of SkM* compared to the SLy4 case.

The overall pion yield is small in the case based on SkM* compared to the cases based on SLy4 and SLy4:L108, in particular at low pion momenta. This can be associated with the relatively small $NN\to N\Delta$ cross sections in nuclear matter in the case of SkM* (see Fig.~\ref{fig:cross}), which is likely a consequence of a relatively weak momentum dependence of the isoscalar nucleon potential in this case (see Fig.~\ref{fig:pot}). However, the overall pion yield will change when the assumed potential for $\Delta$ is varied in the present framework, which nevertheless does not strongly affect the $\pi^-/\pi^+$ ratio (see Sec.~\ref{sec:depDpotiso}).

\subsection{Link from nucleon dynamics to pion observables}
\begin{figure*}
\centering
\includegraphics[scale=0.38]{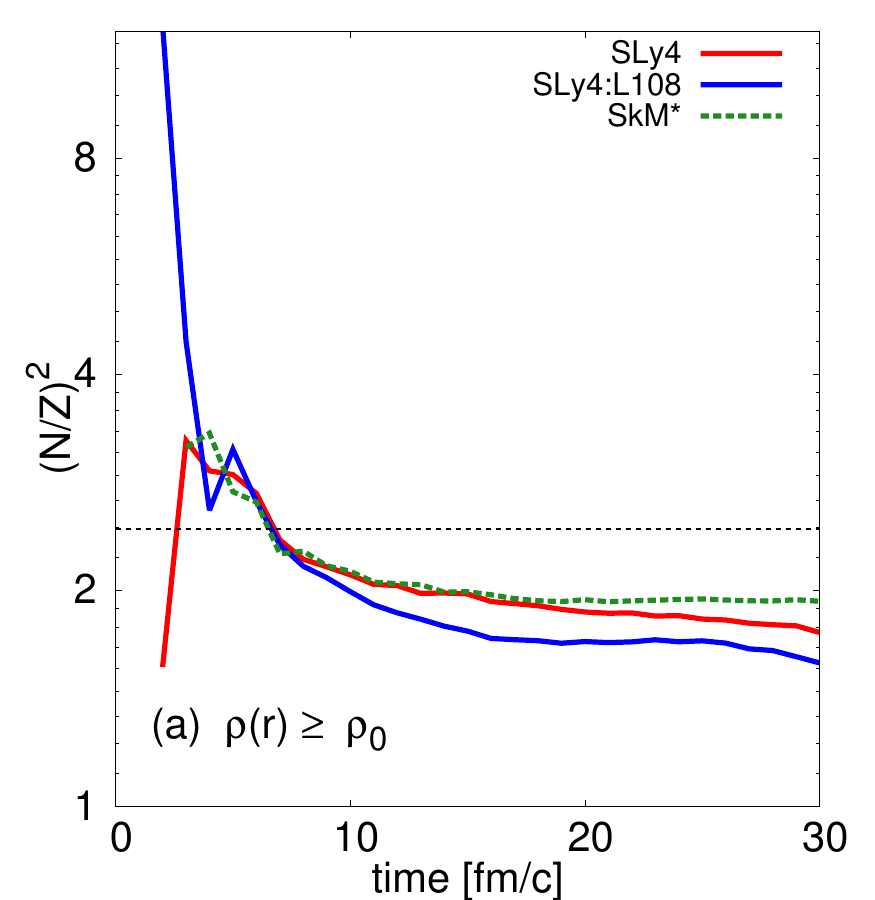}
\includegraphics[scale=0.38]{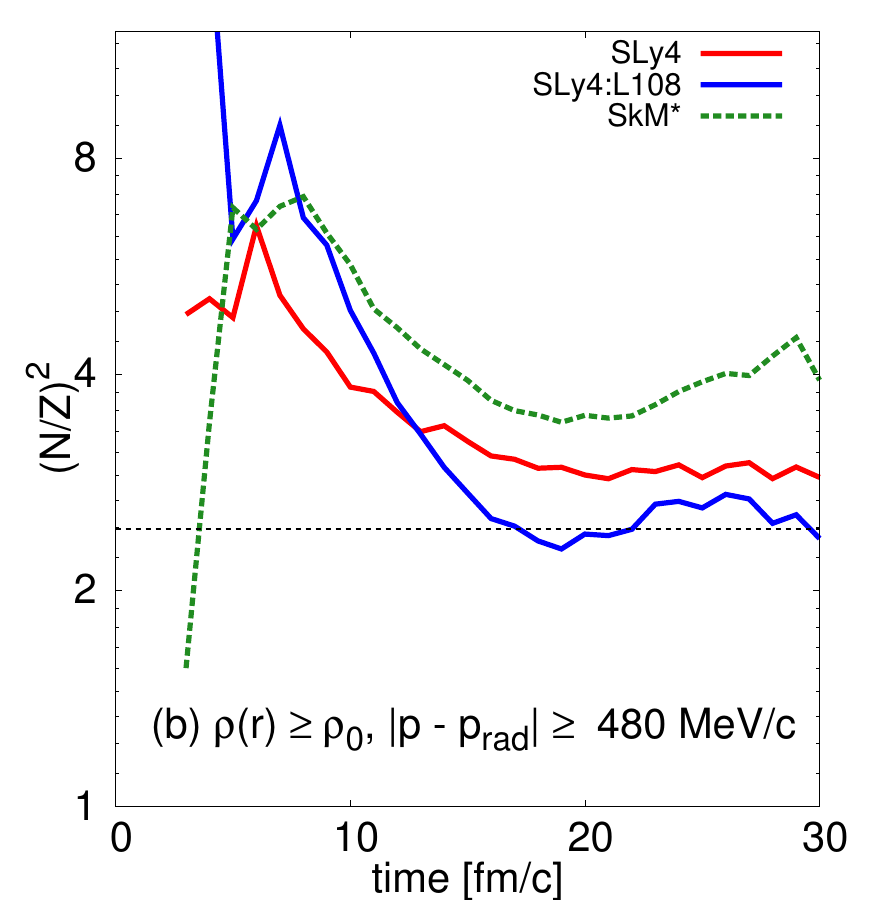}
\includegraphics[scale=0.38]{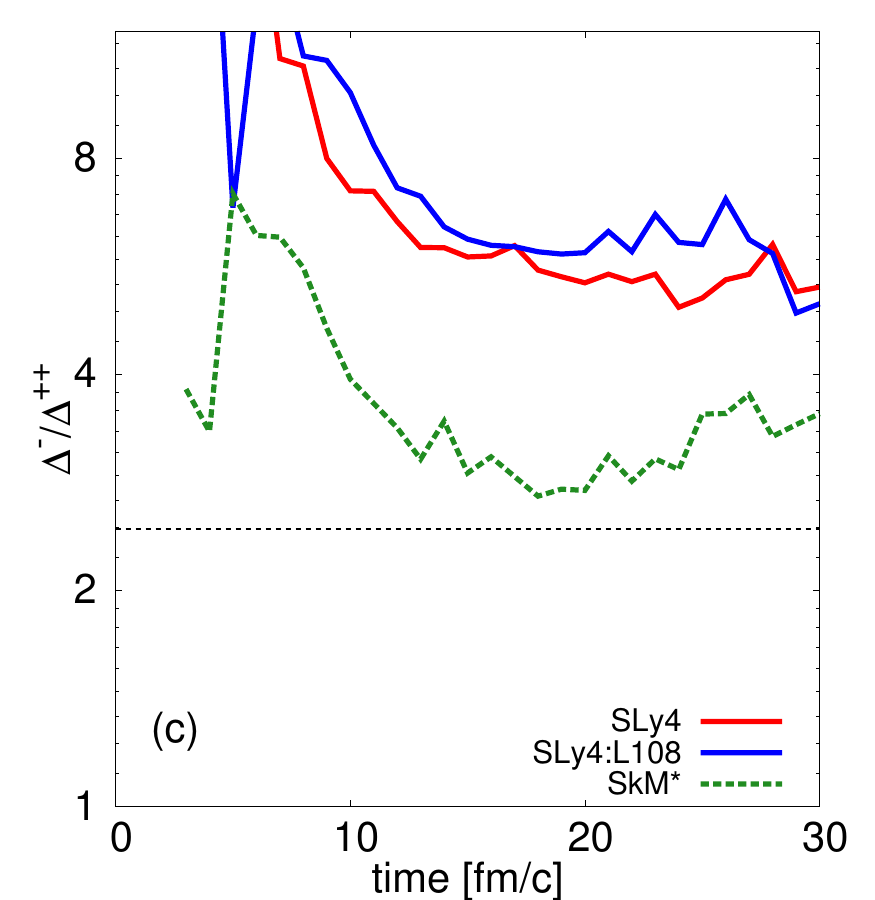}
\caption{ The left two panels (a) and (b) show the time evolution of the squared ratio $(N/Z)^2$ of neutrons and protons, for the three cases of nucleon interaction; panel (a) shows the $(N/Z)^2$ ratio calculated for the nucleons in the high density region ($\rho>\rho_0$), and panel (b) shows the $(N/Z)^2$ ratio for nucleons in the high-density and high-momentum phase-space region (see text). Panel (c) shows the time evolution of the $\Delta^{-}/\Delta^{++}$ production ratio, i.e.\ the ratio of the $nn \rightarrow p \Delta^-$ and $pp \rightarrow n \Delta^{++}$ reaction rates.  The horizontal line in each panel represents the $(N/Z)^2_{\rm sys}$ ratio of the total system. }
\label{fig:NZDelta_ratio}
\end{figure*}

One of the original aims of studying charged pion production in heavy-ion collisions has been to probe the symmetry energy at high densities, expecting that pions originate from energetic nucleon-nucleon collisions occurring in the high density region, as predicted by transport model simulations \cite{bali2002,spirit2021PRL}. For this, it is essential to confirm how the nucleon dynamics in heavy-ion collisions, in particular the neutron-to-proton ratio $N/Z$ in the high density region, is reflected in the final $\pi^-/\pi^+$ ratio, after the processes of $\Delta$ production and others. In the work of Refs.~\cite{ikeno2016,ikeno2016erratum}, we showed that the final $\pi^-/\pi^+$ ratio is indeed correlated to the $N/Z$ ratio in the high-density and high-momentum phase-space region, which the nucleon dynamics determines depending on the high-density symmetry energy. However, this study did not take into account the effects of potentials in the processes related to $\Delta$ and pions.

Here, we repeat the same analysis of Ref.~\cite{ikeno2016}, for the present calculation with potentials in $NN\leftrightarrow N\Delta$ and $\Delta\leftrightarrow N\pi$ processes.  The panel (a) of Fig.~\ref{fig:NZDelta_ratio} shows the time evolution of the squared neutron-to-proton ratio $(N/Z)^2_{\rho>\rho_0}$ in the high density region as a function of time, for the three cases of the nucleon interaction by the three lines. The high density region is defined in each event as the interior of the sphere defined by $\rho(r)>\rho_0$, with $\rho(r)$ being the average density on the sphere of the radius $r$ in the center-of-mass frame of the system. The effect of the symmetry energy is clearly seen by comparing the asy-soft cases (SLy4- and SkM*-based) and the asy-stiff case (SLy4:L108-based) in the time interval $t=15$--$25$ fm/$c$ around which the maximum density is reached.

The panel (b) of Fig.~\ref{fig:NZDelta_ratio} shows the ratio $(N/Z)^2_{\rho>\rho_0,\text{HM}}$ in high-density and high-momentum region, for which the nucleons in the high density region of $\rho(r)>\rho_0$ are further selected by the high momentum condition $|\bm{p} - \bm{p}_{\rm rad}| > p_{\rm cut}$. We take the same condition as in Ref.~\cite{ikeno2016}, i.e., $p_{\rm cut} =480$~MeV/$c$ is chosen and the radial flow $\bm{p}_{\rm rad}=p_{\text{rad}}(r)\bm{r}/r$ is subtracted with $p_{\text{rad}}(r)$ being the radial momentum component averaged for the nucleons on the sphere of the radius $r$. As we have already seen in Refs.~\cite{ikeno2016,ikeno2016erratum}, the ratio $(N/Z)_{\rho>\rho_0,\text{HM}}$ increases compared to $(N/Z)_{\rho>\rho_0}$ when the high momentum region is selected, and the symmetry energy dependence is somewhat enhanced between the SLy4 and SLy4:L108 cases. Furthermore, we can now find a strong effect of the momentum dependence of the neutron and proton potentials ($U_n$ and $U_p$) in the behavior in the SkM* case, where $(N/Z)^2_{\rho>\rho_0,\text{HM}}$ increases most drastically from $(N/Z)^2_{\rho>\rho_0}$ compared to the other cases.  This can be understood from $U_n$ and $U_p$ in the high momentum region of Fig.~\ref{fig:pot}.  Due to a relatively weak momentum dependence of $U_n$, its value at a high momentum in the SkM* case is lower than that in the SLy4 case. Therefore, high-momentum neutrons are favored and thus $(N/Z)^2_{\rho>\rho_0,\text{HM}}$ goes up in the SkM* case compared to the SLy4 case.

Since high-momentum nucleons are responsible to $\Delta$ excitation, we might expect some relation between $(N/Z)^2_{\rho>\rho_0,\text{HM}}$ and the $\Delta$ production by $NN\to N\Delta$. The panel (c) of Fig.~\ref{fig:NZDelta_ratio} shows the ratio between the reaction rates of $nn\to p\Delta^-$ and $pp\to n\Delta^{++}$ (labeled as $\Delta^- / \Delta^{++}$) as a function of time. We recall that $\Delta^-/\Delta^+$ was closely related to $(N/Z)^2_{\rho>\rho_0,\text{HM}}$ when potentials were not taken into account for the $\Delta$ production in Refs.~\cite{ikeno2016,ikeno2016erratum,Ikeno:2019mne}. This is no longer the case when potentials are considered. The $\Delta^-/ \Delta^{++}$ ratio in the SLy4- and SLy4:L108-based cases increases significantly from $(N/Z)^2_{\rho>\rho_0,\text{HM}}$. This increase is the strongest in the SLy4:L108 case. On the other hand, in the SkM*-based case, the $\Delta^-/\Delta^{++}$ ratio becomes lower than $(N/Z)^2_{\rho>\rho_0,\text{HM}}$.  
These drastically different ways of change can be understood again based on the momentum dependence of $U_n$ and $U_p$ shown in Fig.~\ref{fig:pot}, from which we have understood the channel dependence of $NN\to N\Delta$ cross sections in asymmetric nuclear matter (see Sec.~\ref{sec:nnnd}). 
In the $NN \to N \Delta$ reaction, the initial nucleons have high momenta and the final nucleon has a low momentum.  
Considering the $nn \to p \Delta^-$ reaction, 
a high-momentum neutron in the SLy4 and SLy4:L108 cases more favorably turns to a low-momentum proton compared to the SkM* case, because of the difference in the momentum dependence of $U_n$ and $U_p$. 
Next, when the SLy4 and SLy4:L108 cases are compared, a high-momentum neutron in the SLy4:L108 case more favorably turns to a low-momentum proton compared to the SLy4 case, because the difference between neutron and proton potentials at the high density, which is related to the symmetry energy, is larger in the SLy4:L108 case than in the SLy4 case.
Thus, $\Delta^-$ production is favored in the SLy4:L108 case, and consequently the relation between the SLy4 (asy-soft) and SLy4:L108 (asy-stiff) case in $(N/Z)^2$ is inverted in the $\Delta^-/\Delta^{++}$ production ratio.  Namely, the $(N/Z)^2$ ratio with the soft symmetry energy is larger than that with the stiff symmetry energy, while the $\Delta^{-}/\Delta^{++}$ ratio with the stiff symmetry energy is now larger than that with the soft symmetry energy.
The inversion of the symmetry energy effect is also found in other calculations on the pion ratio $\pi^-/\pi^+$~\cite{cozma2016,zhenzhang2017} and the kaon ratio $K^0/K^+$~\cite{ferini2006}
(see, e.g., Ref.~\cite{Colonna:2020euy} for a review).

\begin{figure}
\centering
\includegraphics[width=\columnwidth]{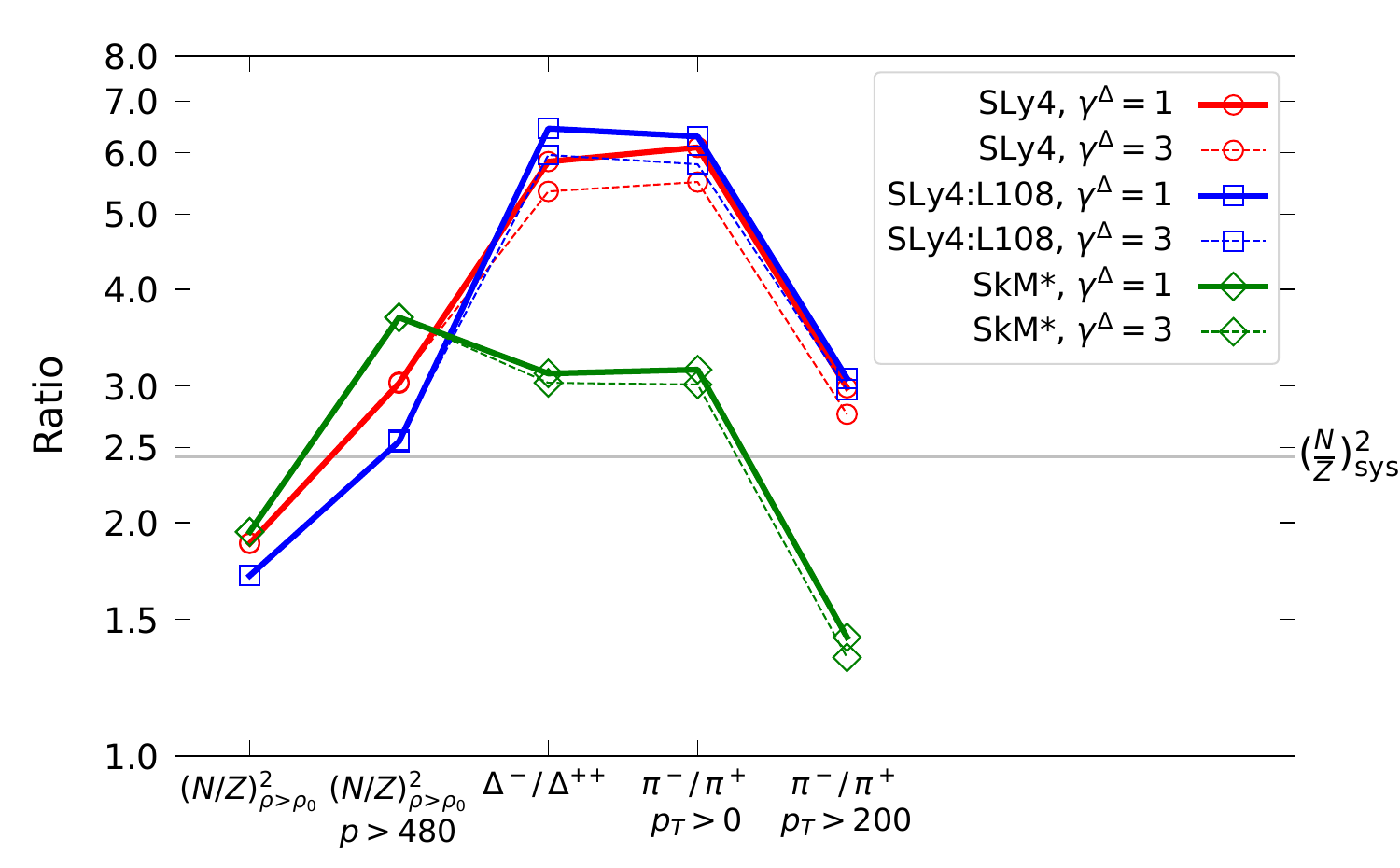}
\caption{ From left to right, the representative nucleon ratios of $(N/Z)^2_{\rho>\rho_0}$ and $(N/Z)^2_{\rho>\rho_0,\text{HM}}$ in the high density region without and with imposing the high momentum condition, respectively, the $\Delta^{-}/\Delta^{++}$ production ratio, the $\pi^-/\pi^+$ ratio from all pions ($p_T> 0$), and the $\pi^-/\pi^+$ ratio of high-momentum pions ($p_T>200$ MeV/$c$).  The results are shown for the three cases of nucleon interaction based on SLy4, SLy4:L108, and SkM*. As for the isovector part of the $\Delta$ potential, two cases are shown for $\gamma^\Delta = 1$ (solid line) and $\gamma^\Delta=3$ (thin dashed line).  The horizontal line represents the $(N/Z)^2_{\rm sys}$ ratio of the total system.  }
\label{fig:ratio}
\end{figure}

The results on the various ratios are concisely summarized in Fig.~\ref{fig:ratio}, which is similar to a figure in Refs.~\cite{ikeno2016,ikeno2016erratum}. In the first and second columns of Fig.~\ref{fig:ratio}, we show a representative $(N/Z)^2$ ratio which is defined in Ref.~\cite{ikeno2016} as
\begin{eqnarray}
(N/Z)^2
  =\frac{ \int_{0}^{\infty} N(t)^{2} dt}{\int_{0}^{\infty} Z(t)^{2} dt},
\label{eq_NZ2}
\end{eqnarray}
where $N(t)$ and $Z(t)$ indicate the numbers of neutrons and protons as functions of time which are selected by the high density condition $\rho>\rho_0$ with or without imposing the high momentum condition $|\bm{p}-\bm{p}_{\text{rad}}|>p_{\text{cut}}$. In the third column of Fig.~\ref{fig:ratio}, we show the representative value of the $\Delta^{-}/\Delta^{++}$ production ratio which is defined in Ref.~\cite{ikeno2016} as
\begin{eqnarray}
 \Delta^{-}/\Delta^{++}=
 \frac{ \int_{0}^{\infty}(nn \rightarrow p \Delta^{-})dt}{\int_{0}^{\infty} (pp \rightarrow n \Delta^{++})dt},
\label{eq_Delta}
\end{eqnarray}
where $(nn \rightarrow p \Delta^{-})$ and $(pp \rightarrow n \Delta^{++})$ indicate the reaction rates of the $\Delta$ production as a function of time. These three representative ratios in Fig.~\ref{fig:ratio} show various effects of the symmetry energy and the momentum dependence of $U_n$ and $U_p$, which we have seen in Fig.~\ref{fig:NZDelta_ratio} and do not repeat here.

In Fig.~\ref{fig:ratio}, we can also find some information on the dependence on the isovector part of the $\Delta$ potential, by comparing two cases of the isospin splitting parameter $\gamma^\Delta=1$ (solid line) and $\gamma^\Delta=3$ (thin dashed line). In the present calculation, the splitting plays a minor but non-negligible role in the $\Delta$ production.

As for the pion ratios, the fourth column shows the $\pi^-/\pi^+$ ratio calculated from all pions ($p_T>0$), while the fifth column shows that from the high-momentum pions selected by the transverse momentum $p_T> 200$ MeV/$c$, which corresponds to the region that Ref.~\cite{spirit2021PRL} used to extract information on the symmetry energy from the S$\pi$RIT experimental data. We include here all pions emitted to both forward and backward angles. The $\pi^-/\pi^+$ ratio from all pions ($p_T>0$) is almost identical to the $\Delta^-/\Delta^{++}$ production ratio in all the cases, except the $\pi^-/\pi^+$ ratio in the asy-soft case (SLy4) slightly increases from the $\Delta^-/\Delta^{++}$ ratio. The reduction of the $\pi^-/\pi^+$ ratio for $p_T>200$ MeV/$c$ is due to the effect of the Coulomb force acting on charged pions, which is known to be well under control in transport models \cite{xu2023}. Thus, the final pion ratio is rather simply related to the $NN\to N\Delta$ process. 

In summary, the impact of a change of nuclear symmetry energy (SLy4- vs SLy4:L108-based nucleon interaction) on the pion ratio is not very large and it is a consequence of different effects in the nucleon dynamics and in the $NN\to N\Delta$ process which act in opposite directions. The impact of the isospin splitting of the $\Delta$ potential can be of the same order of that of nuclear symmetry energy. Much larger is the impact of a change of the momentum dependence of the neutron and proton potentials (SLy4- vs SkM*-based). This is also a consequence of the effects in the nucleon dynamics and in the $NN\to N\Delta$ process acting in opposite directions; however, the effect in $NN\to N\Delta$ is much larger.

\subsection{The effect of the in-medium $\Delta$ spreading width \label{sec:SpreadGam}}

\begin{figure}
\centering
\includegraphics[width=\columnwidth]{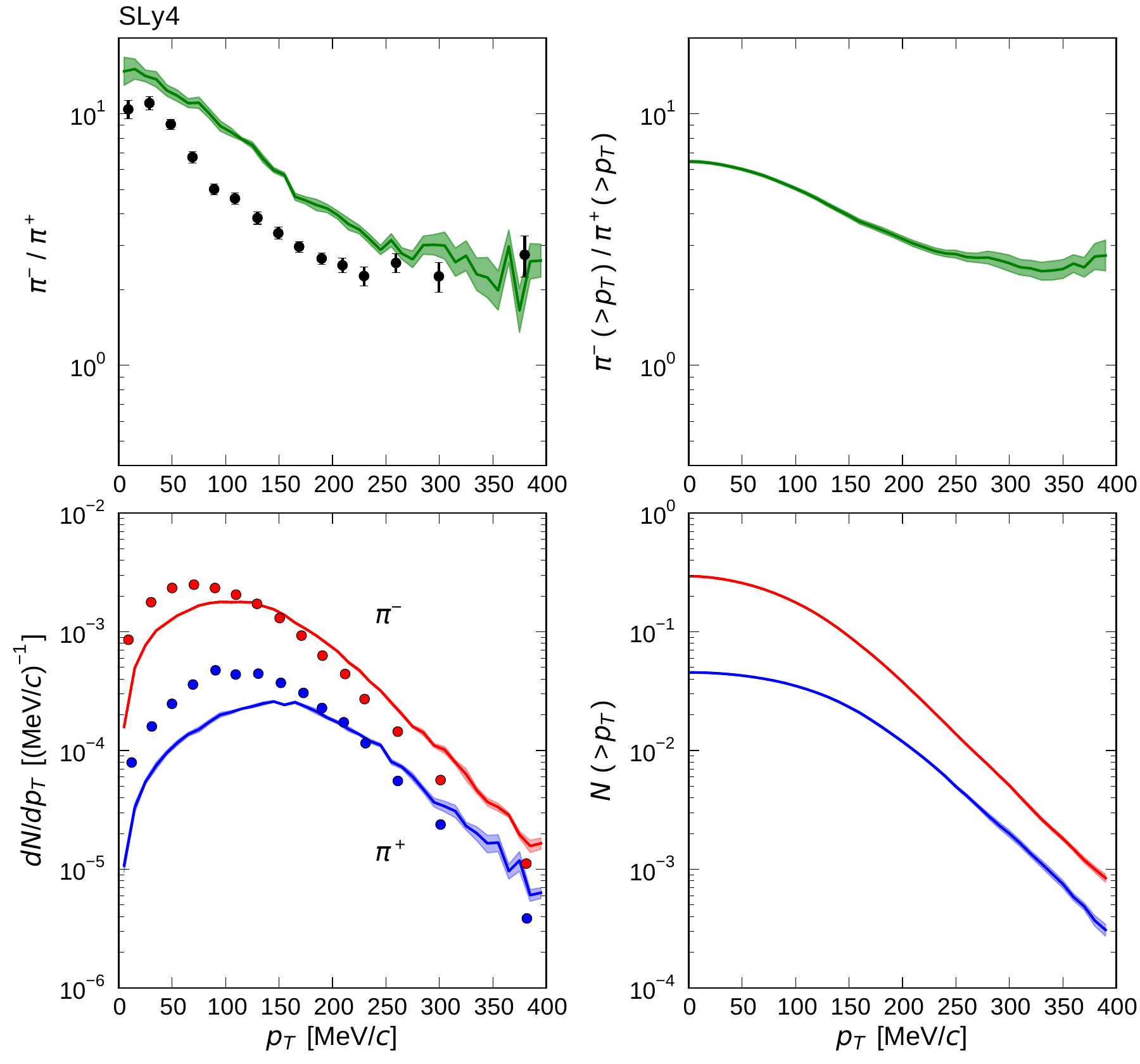}
\caption{The same as Fig.~\ref{fig:pionspectra-sly4} except the in-medium spreading width of $\Delta$ is turned off by setting $\Gamma_{\text{sp}}^\Delta=0$. 
The repulsive terms in the isoscalar part of the $\Delta$ potential are included by $\alpha_\rho^\Delta=15$ MeV and $\alpha_\tau^\Delta=15$ MeV as Fig.~\ref{fig:pionspectra-sly4}.
}
\label{fig:pionspectra-sly4_exgam0}
\centering
\includegraphics[width=\columnwidth]{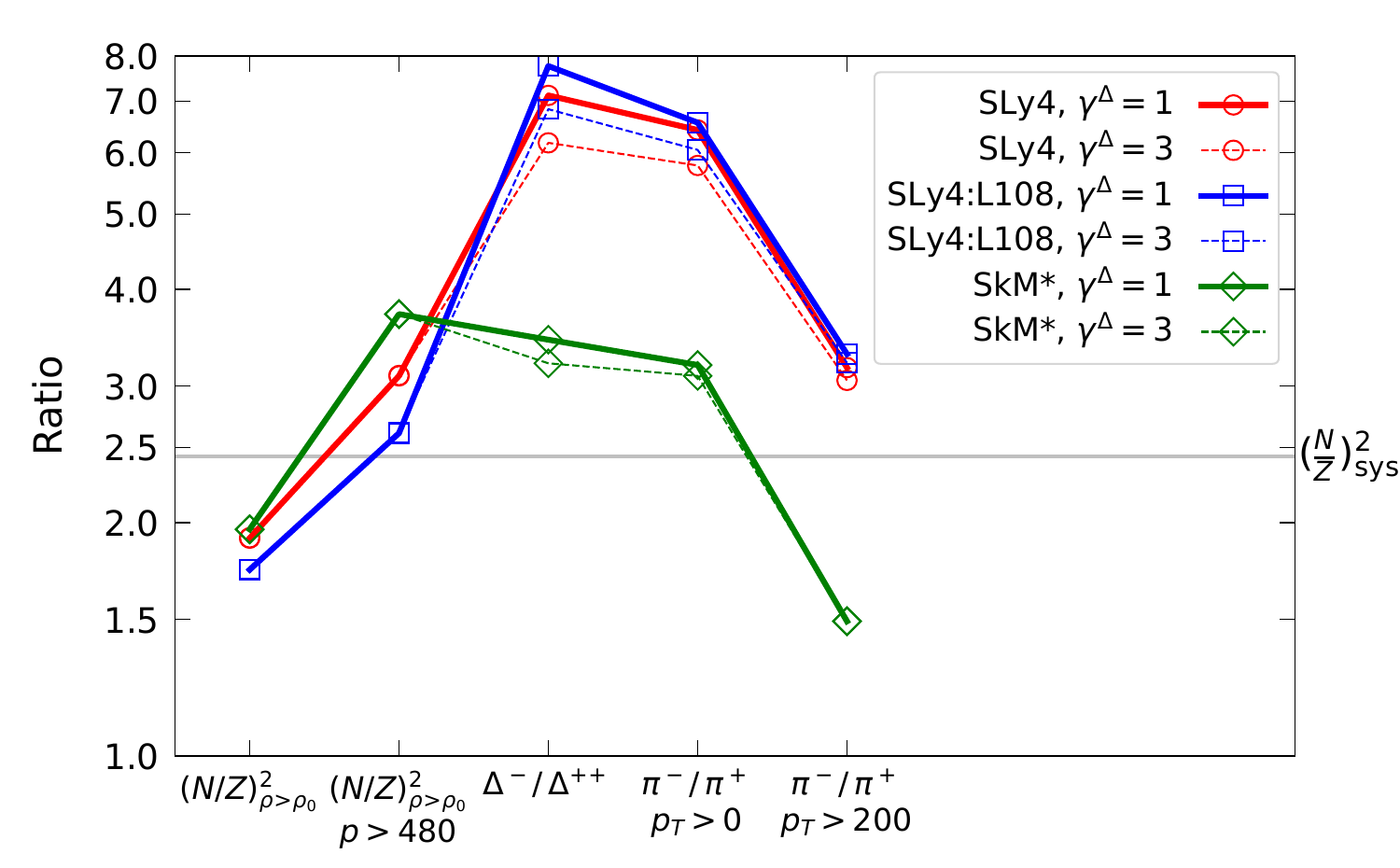}
\caption{The same as Fig.~\ref{fig:ratio} except the in-medium spreading width of $\Delta$ is turned off by setting $\Gamma_{\text{sp}}^\Delta=0$.
The repulsive terms in the isoscalar part of the $\Delta$ potential are included by $\alpha_\rho^\Delta=15$ MeV and $\alpha_\tau^\Delta=15$ MeV as Fig.~\ref{fig:ratio}.
}
\label{fig:ratios_exgam0}
\end{figure}

So far, we have calculated the pion production with the default option ($\alpha_\rho^\Delta=15$~MeV, $\alpha_\tau^\Delta=15$~MeV, and $\Gamma_{\text{sp}}^\Delta=60$~MeV) which added the repulsive terms in the isoscalar part of the $\Delta$ potential and the spreading width $\Gamma_{\text{sp}}^\Delta$.
Here, in order to see the effect of the spreading width of $\Delta$ in the medium [see Eq.~\eqref{eq:TotalGam_Delta}],
we show in Fig.~\ref{fig:pionspectra-sly4_exgam0} the pion spectra when the spreading width is turned off ($\Gamma_{\text{sp}}^\Delta=0$).
By comparing the lower left panel of Fig.~\ref{fig:pionspectra-sly4_exgam0} with that of Fig.~\ref{fig:pionspectra-sly4}, we can see clearly that the spreading width affects only the low momentum part of the $\pi^-$ and $\pi^+$ spectra.
Namely, the spreading width works to increase the pion yield in the low momentum region.

In spite of the change of the $\pi^-$ and $\pi^+$ spectra, the $\pi^-/\pi^+$ ratio of the spectra, shown in the upper panels of Fig.~\ref{fig:pionspectra-sly4_exgam0}, is not affected much when the spreading width is turned off. Thus, the pion ratio is almost free from the uncertainties in the in-medium spreading width.

Figure.~\ref{fig:ratios_exgam0} shows the various ratios when the spreading width is turned off by $\Gamma_{\text{sp}}^\Delta=0$. The $\pi^-/\pi^+$ ratios here are quantitatively similar to those in Fig.~\ref{fig:ratio} where the spreading width parameter was $\Gamma_{\text{sp}}^\Delta=60$~MeV. On the other hand, the $\Delta^-/\Delta^{++}$ production ratios become larger than those in Fig.~\ref{fig:ratio}.
We can also see that the effect of the symmetry energy (SLy4- vs SLy4:L108-based) is small, and the effect of the difference in the momentum dependence of $U_n$ and $U_p$ (SLy4- vs SkM*-based) is much more significant. These trends are the same as shown in Fig.~\ref{fig:ratio}.

\subsection{The effect of the isoscalar part of the $\Delta$ potentials\label{sec:depDpotiso}}
Finally, we test the robustness of the above results against the uncertainties in the isoscalar part of the $\Delta$ potential, for which we added repulsive terms with parameters $\alpha_\rho^\Delta$ and $\alpha_\tau^\Delta$, compared to the SkM*-based nucleon potential [see Eq.~\eqref{eq:Uis_Delta}]. 

\begin{figure}
\centering
\includegraphics[width=\columnwidth]{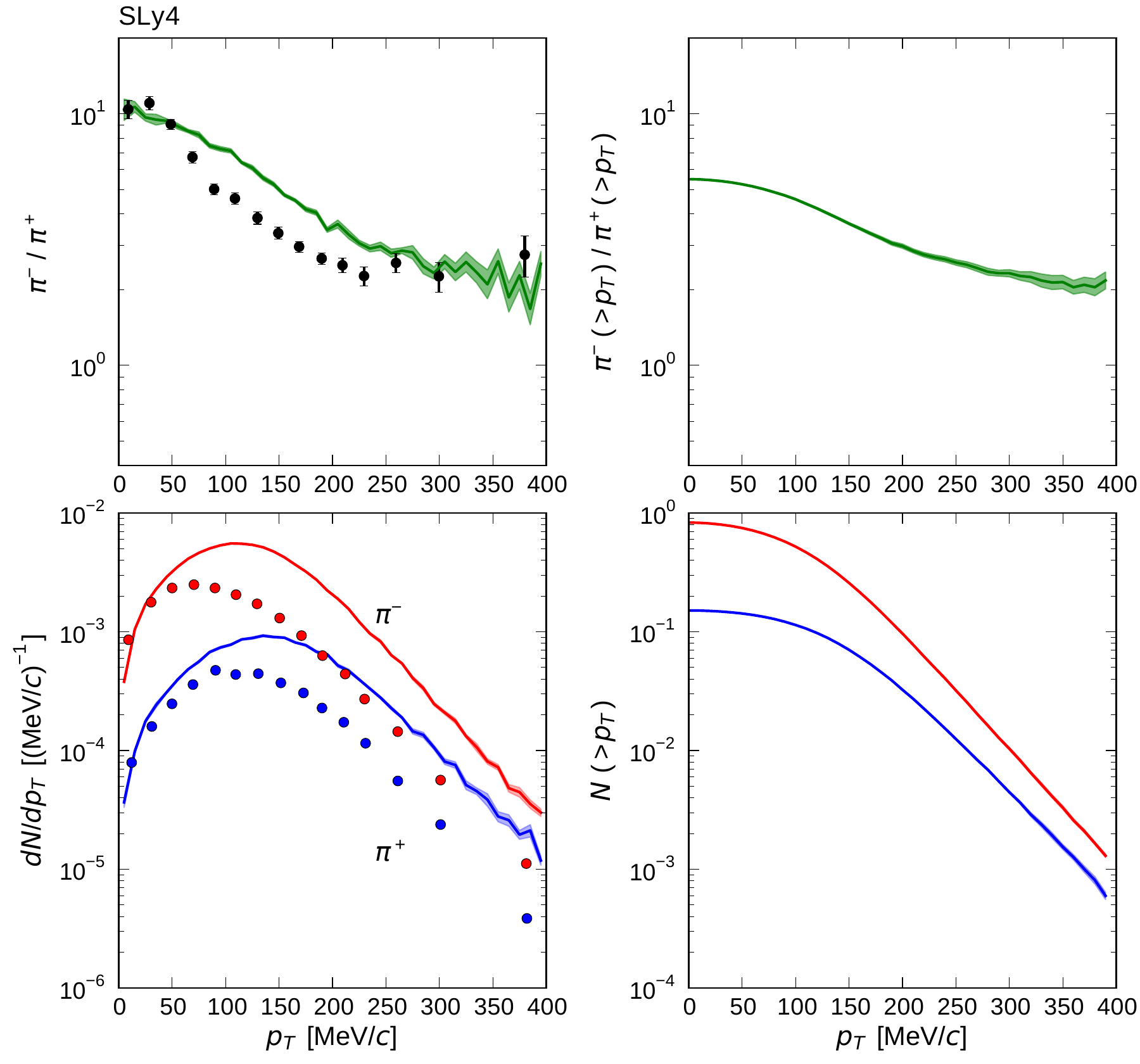}
\caption{The same as Fig.~\ref{fig:pionspectra-sly4_exgam0} except the repulsive terms in the isoscalar part of the $\Delta$ potential is turned off by setting $\alpha_\rho^\Delta = 0$, $\alpha_\tau^\Delta=0$ and $\Gamma_{\text{sp}}^\Delta=0$.
}
\label{fig:pionspectra-sly4_alphatau00}
\centering
\includegraphics[width=\columnwidth]{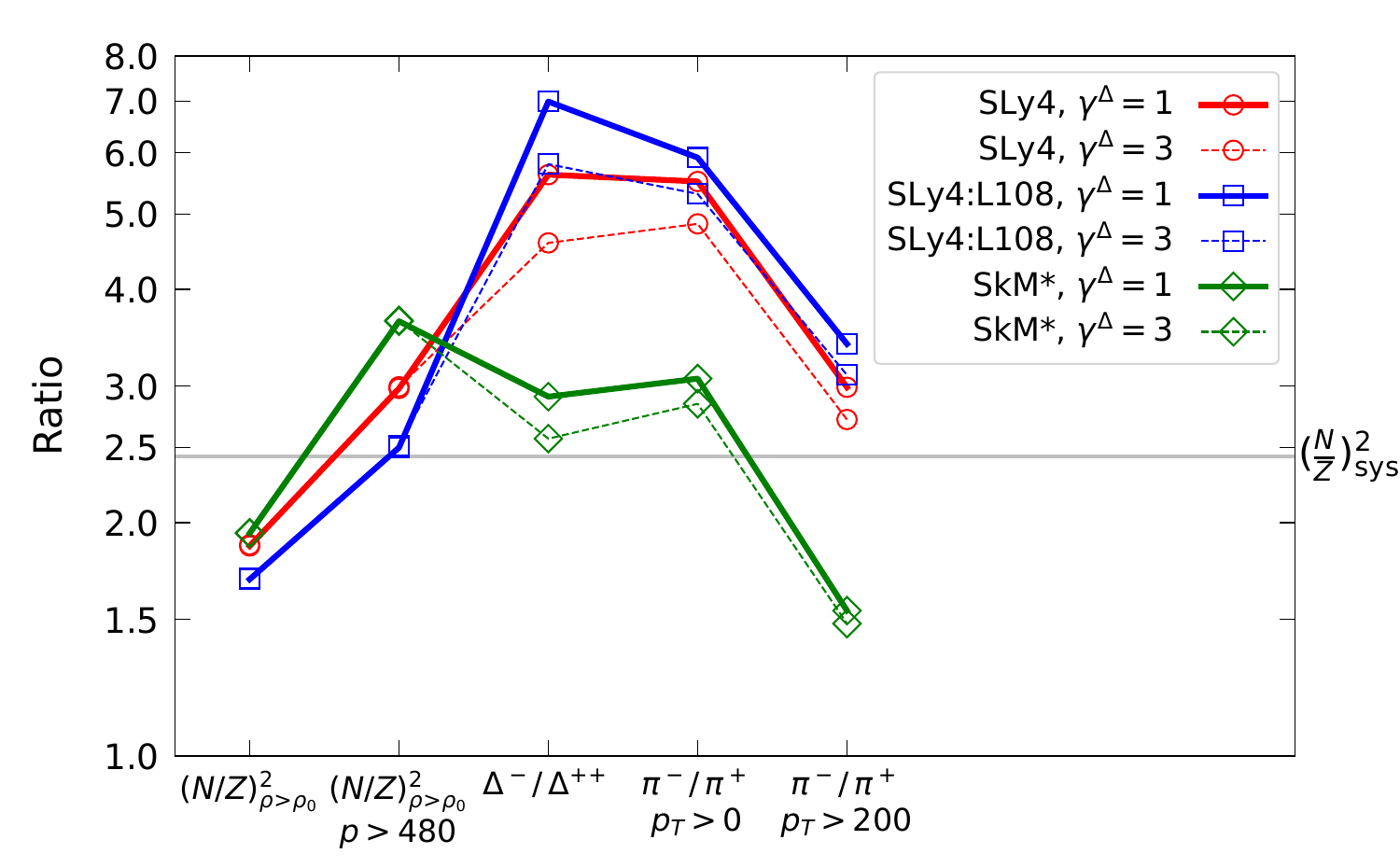}
\caption{The same as Fig.~\ref{fig:ratios_exgam0} except the repulsive terms in the isoscalar part of the $\Delta$ potential is turned off by setting $\alpha_\rho^\Delta = 0$, $\alpha_\tau^\Delta=0$ and $\Gamma_{\text{sp}}^\Delta=0$.
}
\label{fig:ratio_alphatau00}
\end{figure}

When these options are turned off by setting $\alpha_\rho^\Delta=0$ and $\alpha_\tau^\Delta=0$ (and we also set $\Gamma_{\text{sp}}^\Delta=0$ in this subsection as well as in Subsec.~\ref{sec:SpreadGam}), 
the pion yield is overestimated as we can see in the lower left panel of Fig.~\ref{fig:pionspectra-sly4_alphatau00} in comparison with that of Fig.~\ref{fig:pionspectra-sly4_exgam0}. This is naturally understood as a consequence of turning off the repulsive terms in the $\Delta$ potential. 
Even in this case, the $\pi^-/\pi^+$ ratio of the spectra, shown in the upper panels of Fig.~\ref{fig:pionspectra-sly4_alphatau00}, is not affected much by changing the parameters $\alpha_\rho^\Delta$ and $\alpha_\tau^\Delta$.
Therefore, the result here also suggests that the pion ratio is not affected strongly by the uncertainties in the isoscalar $\Delta$ potential.

Figure~\ref{fig:ratio_alphatau00} shows the various ratios when the repulsive terms are turned off by $\alpha_\rho^\Delta=0$ and $\alpha_\tau^\Delta=0$ (and $\Gamma_{\text{sp}}^\Delta=0$). The $\Delta^-/\Delta^{++}$ production ratio and the $\pi^-/\pi^+$ ratio here are quantitatively similar to those in Fig.~\ref{fig:ratios_exgam0} where the repulsive terms were taken into account. However, the effect of the symmetry energy (SLy4- vs SLy4:L108-based) is now stronger, i.e., we find a stronger inversion of the symmetry energy effect from the $(N/Z)^2_{\rho>\rho_0,\text{HM}}$ ratio to the $\Delta^-/\Delta^{++}$ production ratio when the repulsive terms in the $\Delta$ potential is turned off. The effect of the isospin splitting of the $\Delta$ potential ($\gamma^\Delta=1$ vs $\gamma^\Delta=3$) is also stronger in Fig.~\ref{fig:ratio_alphatau00} compared to that in Fig.~\ref{fig:ratios_exgam0}. On the other hand, the effect of the difference in the momentum dependence of $U_n$ and $U_p$ (SLy4- vs SkM*-based) is always the most significant, which is not affected by the uncertainties in the isoscalar part of the $\Delta$ potential and the in-medium spreading width of $\Delta$.

\section{Summary\label{sec:summary}}

We investigated the production of $\Delta$ resonances and pions in ${}^{132}\mathrm{Sn}+{}^{124}\mathrm{Sn}$ collisions at $E/A=270$ MeV/nucleon within the AMD+sJAM model, in which the collision term takes into account the momentum-dependent mean-field potentials with strict conservation of energy and momentum. In the newly developed part sJAM of the model, the potentials for the particles in the initial and final states of a process is treated in the form of the scalar and vector self-energies for each species of particles, and the potentials affect the phase space factor for the final state and the flux factor for the initial state in a natural way.  In particular, the cross section for $NN\to N\Delta$ depends on the isospin channel when the neutron and proton potentials are different in isospin-asymmetric environment. The mass distribution or the spectral function of the $\Delta$ resonance is also determined by the potentials through the potential dependence of the $\Delta\to N\pi$ width.

In particular, we focused on the effect of the different momentum dependence between the neutron and proton potentials. When the neutron potential has a strong momentum dependence compared to the proton potential in a neutron-rich environment ($m_n^* < m_p^*$ in the SLy4-based case), the process $nn\to p\Delta^-$ that converts two high-momentum neutrons to a low momentum proton is favored compared to the $pp\to n\Delta^{++}$ process. The tendency is opposite when the neutron potential has a weak momentum dependence compared to the proton potential ($m_n^* > m_p^*$ in the SkM*-based case). The result of the AMD+sJAM simulations shows that this effect of the momentum dependence appears very clearly in the $\Delta$ production and consequently the $\pi^-/\pi^+$ ratio is an observable that is very sensitive to the momentum dependence of the neutron and proton potentials. The case with a strong momentum dependence of neutrons compared to protons is more consistent with the S$\pi$RIT data than the opposite case.

We also investigated the effects of other ingredients. The symmetry energy $L$ dependence (SLy4 vs SLy4:L108) was found to have a relatively small effect on the pion ratio compared to the effect of the momentum dependence of the nucleon potentials. We carefully traced a link from the nucleon dynamics to the pion observable through the $\Delta$ production, and the symmetry energy effect in the neutron-proton ratio $(N/Z)$ was found to be reversed in that of $\Delta$ production rate ($\Delta^{-}/\Delta^{++}$). 
We confirmed that the conclusions remain the same even if the isoscalar and isovector parts of the $\Delta$ potential and the in-medium $\Delta$ spreading width are changed.

We have found that the in-medium $\Delta$ spreading width, due to the collisions of $\Delta N\to NN$ and $\Delta N\to\Delta N$,
affects only the low-momentum part of the pion spectrum.
It is known by other transport model calculations~\cite{spirit2021PRL,cozma2021} that the low-energy part is also sensitive to the pion potential which the present calculation ignored. Therefore, it is desirable to obtain a full understanding of the low energy pion emission in the future by considering both the $\Delta$ spreading width and the pion potential.
On the other hand, 
the high-momentum pions may be suitable to extract physics information from experiments as claimed in Refs.~\cite{spirit2021PRL,cozma2021}.
We will check this point in our future work
by investigating the role of the pion potential which should be considered consistently for both the energy conservation and the rates in the $\Delta \leftrightarrow N \pi$ processes, as well as for the propagation of pions.

\section*{Acknowledgments}
N.~I. would like to thank Che Ming Ko for valuable discussions and encouragement, Zhen Zhang for the practical information on the Hama potential, Betty Tsang and Tommy Tsang for information on the experimental data of S$\pi$RIT, 
and Eulogio Oset for discussing the $\Delta$ potential. We thank Yasushi Nara for the useful information on the formula of the $\Delta$ decay width.
The computation was carried out at the HOKUSAI supercomputer system of RIKEN.
This work was supported by JSPS KAKENHI Grant Numbers JP17K05432, JP19K14709, JP21KK0244, and JP21K03528.

\bibliography{ref_nucl}

\end{document}